\documentclass[a4paper,fleqn,usenatbib]{mnras}

\usepackage{newtxtext,newtxmath}

\usepackage[T1]{fontenc}
\usepackage{ae,aecompl}

\usepackage{graphicx}	
\usepackage{amsmath}	

\title[3D stellar evolution] {3D stellar evolution: hydrodynamic simulations of a complete burning phase in a massive star}

\author[Rizzuti et al.]{F. Rizzuti$^{1}$\thanks{E-mail: f.rizzuti@keele.ac.uk}, R. Hirschi$^{1,2}$, W. D. Arnett$^{3}$, C. Georgy$^{4}$, C. Meakin$^{5}$, A. StJ. Murphy$^{6}$, T. Rauscher$^{7,8}$ \newauthor and V. Varma$^{1}$
\\
$^{1}$Astrophysics Group, Lennard-Jones Laboratories, Keele University, Keele ST5 5BG, UK\\
$^{2}$Kavli IPMU (WPI), University of Tokyo, 5-1-5 Kashiwanoha, Kashiwa 277-8583, Japan\\
$^{3}$Steward Observatory, University of Arizona, 933 N. Cherry Avenue, Tucson AZ 85721, USA\\
$^{4}$Geneva Observatory, Geneva University, CH-1290 Sauverny, Switzerland\\
$^{5}$Pasadena Consulting Group, 1075 N Mar Vista Ave, Pasadena, CA 91104 USA\\
$^{6}$School of Physics and Astronomy, University of Edinburgh, Edinburgh EH9 3FD, UK\\
$^{7}$Department of Physics, University of Basel, Switzerland\\
$^{8}$Centre for Astrophysics Research, University of Hertfordshire, UK
}

\date{Accepted 2023 May 22. Received 2023 May 17; in original form 2023 March 25}

\pubyear{2023}

\begin{document}
\label{firstpage}
\pagerange{\pageref{firstpage}--\pageref{lastpage}}
\maketitle

\begin{abstract}
Our knowledge of stellar evolution is driven by one-dimensional (1D) simulations. 1D models, however, are severely limited by uncertainties on the exact behaviour of many multi-dimensional phenomena occurring inside stars, affecting their structure and evolution. Recent advances in computing resources have allowed small sections of a star to be reproduced with multi-D hydrodynamic models, with an unprecedented degree of detail and realism. In this work, we present a set of 3D simulations of a convective neon-burning shell in a 20 M$_\odot$ star run for the first time continuously from its early development through to complete fuel exhaustion, using unaltered input conditions from a 321D-guided 1D stellar model. These simulations help answer some open questions in stellar physics. In particular, they show that convective regions do not grow indefinitely due to entrainment of fresh material, but fuel consumption prevails over entrainment, so when fuel is exhausted convection also starts decaying. Our results show convergence between the multi-D simulations and the new 321D-guided 1D model, concerning the amount of convective boundary mixing to include in stellar models. The size of the convective zones in a star strongly affects its structure and evolution, thus revising their modelling in 1D will have important implications for the life and fate of stars. This will thus affect theoretical predictions related to nucleosynthesis, supernova explosions and compact remnants.
\end{abstract}




\begin{figure*}
\centering
\footnotesize
\includegraphics[trim={0.33cm -0.25cm 1.25cm 0cm},clip,width=0.58\textwidth]{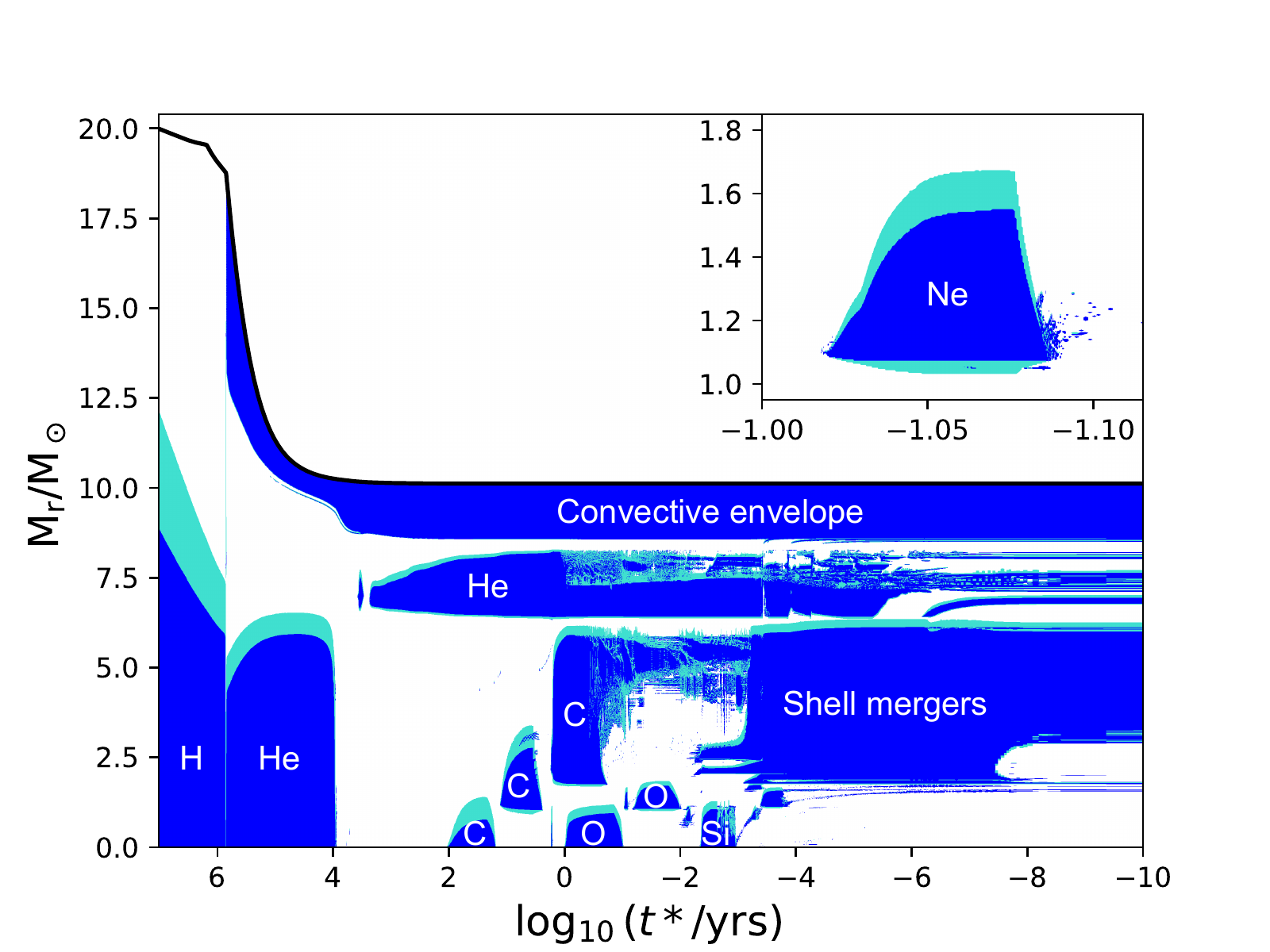}\includegraphics[trim={0.25cm 0cm 0cm 0cm},clip,width=0.45\textwidth]{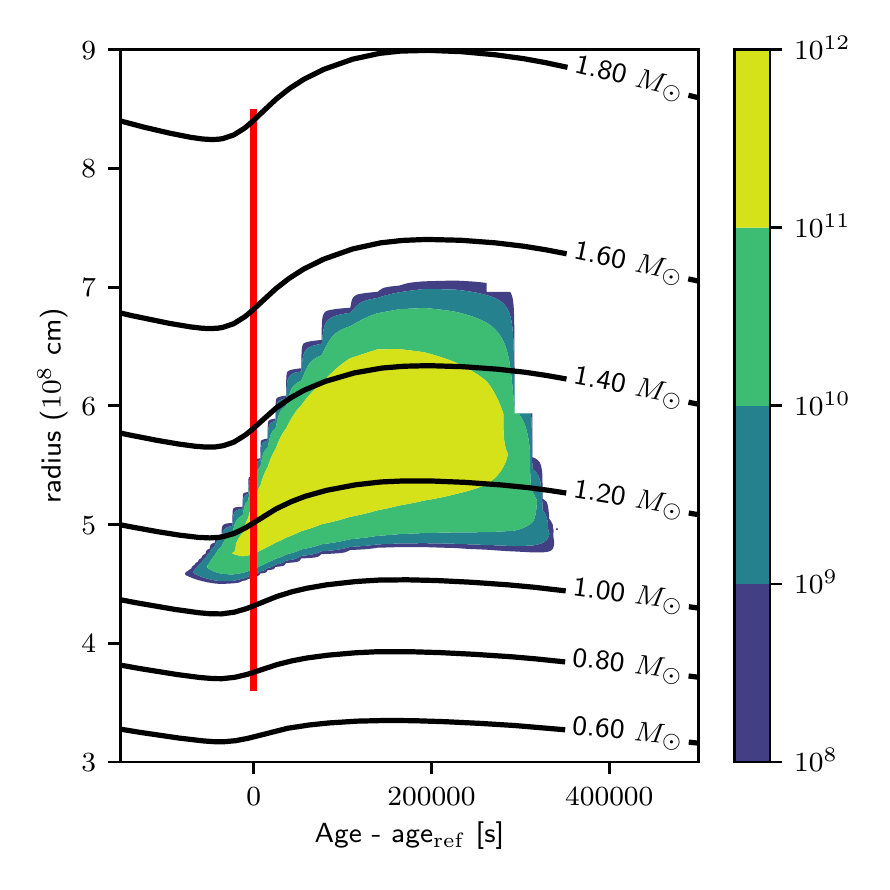}\put(-374,25){\color{red}\Huge\boldmath$\uparrow$}
\caption{{\it Left}: Structure evolution diagram of the 20 M$_\odot$ 1D \texttt{MESA} input stellar model as a function of the time left until the predicted collapse of the star (in years, log scale). In blue the convective zones, in green the CBM zones. The red arrow indicates the neon-burning shell the 3D simulations were started from. Top right corner is a zoom-in on the neon-burning shell. {\it Right:} Zoom-in on the model used as initial conditions for the 3D simulations. The horizontal axis is the time in seconds relative to the start of the 3D simulations. The vertical axis is the radius in units of 10$^{8}$ cm. In colour scale, the squared convective velocity. Isomass contours are shown as black lines. The lines show that the shell undergoes significant expansion during the Ne-burning phase. The vertical red bar indicates the start time and radial extent of the hydrodynamic simulations.}\label{fig:kip}
\end{figure*}

\section{Introduction}
To have good understanding and reliable predictions of the evolution and properties of stars, researchers have developed stellar evolution models that can follow the life cycle of an entire star from birth to death \citep[e.g.][]{2002ApJ...567..532H,Paxton_2011,2012A&A...537A.146E}. However, there are many physical processes to take into account that can affect the evolution and fate of a star. To make the computation possible and affordable, stellar models rely on simplifications and assumptions, one of the most important of which is spherical symmetry. An equally important though less easily envisages simplification is the treatment of convection in stars with mixing length theory \citep{1958ZA.....46..108B}. \\
To improve and refine these prescriptions, sections of a star may be studied in more detail employing multi-dimensional hydrodynamic models, that can realistically reproduce complex physical processes without using any of the prescriptions of the 1D case \citep[e.g.][]{1998ApJ...494..680J, 2009ARA&A..47..481A, 2015ApJ...807L..31L, 2016Janka, 2021Natur.589...29B}. Unfortunately, these simulations are computationally very expensive, requiring millions of core-hours to simulate hours of stellar evolution, so reproducing the evolutionary timescales has so far been inaccessible. Hydrodynamic simulations of stellar interiors mostly focus on reproducing turbulent convection and mixing between layers (known as ``convective boundary mixing'', CBM), resulting in the growth over time of convective zones due to entrainment of material from the stable regions into the convective ones. This is particularly impactful for the late phases of massive stars, where convective speeds are high and entrainment is vigorous, affecting the structure of the star and the end of its life. The very last phases before the onset of iron core-collapse are often studied with multi-D hydrodynamic supernova simulations \citep{2015ApJ...808L..21C}, which recently started to take into account also the effects of rotation \citep{10.1093/mnrasl/slab067, 2022MNRAS.509..818M} and magnetic fields \citep{10.1093/mnras/stab883} on the convective motions. \\
CBM is important also for the convective cores of main-sequence stars, with implications for the stellar lifetime and type of remnant \citep{2020MNRAS.496.1967K}, but here the convective velocities are much smaller and it is difficult for hydrodynamic models to simulate significant time ranges without boosting the rate of energy release to accelerate the simulations (\citealt{2007ApJ...667..448M}; see also the study in \citealt{2023B}), as is also often done for later phases \citep{2017MNRAS.471..279C, 2021A&A...653A..55H}.\\
To validate results, it is crucial to investigate stellar convection in hydrodynamic simulations that do not modify the energy release. Previous studies of 3D stellar simulations do alter the initial conditions introducing some energy boosting, with some exceptions such as the very late oxygen and silicon convective shells run for short timescales \citep{2015Couch,2017MNRAS.472..491M,2019ApJ...881...16Y}. However \cite{2022R}, with their analysis of CBM in a nominal-luminosity neon shell, revealed the presence of strong entrainment in the convective zone, confirming it is not an effect of the boosting in luminosity but a natural result of the convectively unstable structure predicted by the 1D stellar models. This shows the importance of correctly implementing entrainment also in 1D models, to ensure consistency and agreement between results. We call  this approach ``321D-guided'', referring to 1D stellar models that have been improved using the results of 3D hydrodynamic simulations.\\
It is worth noting though that the nominal-luminosity simulation of \cite{2022R} has been run only up to 3000 seconds, which while long enough for studying CBM with good statistics, is too short to give any information on the evolution of the burning shell towards fuel exhaustion. Until now, multi-D studies have not investigated the interplay between entrainment and fuel exhaustion at the end of a burning phase, leaving many questions about the fate of convection unanswered. Specifically, does convection stop when the fuel is exhausted, or does the mixing of fresh fuel from entrained material extend the convective growth indefinitely? This issue, on which 1D models cannot give any information, has a pivotal role for correctly predicting the later evolution of the star, in particular the pre-SN structure and the type of remnant. \\
We organize the paper as follows: in Section~\ref{sec:2}, we present the initial conditions and general setup of the hydrodynamic simulations. In Section~\ref{sec:3}, we analyse the results from the sets of simulations. Finally, we discuss results and draw conclusions in Section~\ref{sec:4}.

\begin{table*}
\centering
\footnotesize
\caption{Properties of the 3D hydrodynamic simulations presented in this study: model name; resolution $N_{r\theta\phi}$; boosting factor of the driving luminosity $\epsilon$; starting $t_\text{start}$ and ending $t_\text{end}$ time of the simulation; time spent in the entrainment regime $\tau_\text{entr}$; convective turnover time $\tau_\text{conv}$; number of convective turnovers simulated in the entrainment regime $n_\text{conv}$; root-mean-square convective velocity $v_\text{rms}$; sonic Mach number \textit{Ma}; number of CPU core-hours required to run the simulation.}\label{tab:1}
\begin{tabular}{lccccccccccc}
\hline
\hline
\texttt{name}&$N_{r\theta\phi}$&$\epsilon$&$t_\text{start}$ (s)&$t_\text{end}$ (s)&$\tau_\text{entr}$ (s)&$\tau_\text{conv}$ (s)&$n_\text{conv}$&$v_\text{rms}$ (cm s$^{-1}$)&\textit{Ma }$(10^{-2})$&Mcore-h\\
\hline
\texttt{r256e1}&$256\times128^2$&1&0&60000&15000&155&96&3.29 $\times$ 10$^6$&0.83&2.08\\
\texttt{r256e5}&$256\times128^2$&5&0&29000&1500&59&25&6.55 $\times$ 10$^6$&1.76&0.89\\
\texttt{r256e10}&$256\times128^2$&10&0&19000&800&50&16&8.06 $\times$ 10$^6$&2.15&0.60\\
\texttt{r256e50}&$256\times128^2$&50&0&30000&150&30&5&1.31 $\times$ 10$^7$&3.48&0.96\\
\texttt{r512e1}&$512\times256^2$&1&16000&19000&3000&136&22&3.83 $\times$ 10$^6$&0.99&1.66\\
\texttt{r512e5}&$512\times256^2$&5&0&2000&1500&59&25&6.65 $\times$ 10$^6$&1.80&0.80\\
\texttt{r512e10}&$512\times256^2$&10&0&1000&800&49&16&8.28 $\times$ 10$^6$&2.23&0.50\\
\texttt{r512e50}&$512\times256^2$&50&0&490&150&30&5&1.34 $\times$ 10$^7$&3.61&0.20\\
\texttt{r1024e1}&$1024\times512^2\enspace$&1&10000&10400&400&127&3&3.26 $\times$ 10$^6$&0.84&2.88\\
\texttt{r2048e1}&$2048\times1024^2$&1&10010&10030&20&113&0&3.85 $\times$ 10$^6$&0.99&2.02\\
\hline
\end{tabular}
\end{table*}

\section{Methods}\label{sec:2}
\subsection{Initial conditions from a 1D stellar model}
To run a stellar hydrodynamic simulation, initial conditions need to be assumed from a 1D stellar evolution model that simulates the entire lifetime of the star, so that the realism of the simulated environment can be ensured. For this purpose, we used the \texttt{MESA} stellar evolution code \citep{Paxton_2011, Paxton_2013, Paxton_2018, Paxton_2019} to model the evolution of a 20 M$_\odot$ star of solar metallicity ($Z=0.014$ using the relative abundances of \citealt{2009ARA&A..47..481A}) from the pre-main-sequence until core-collapse. Mass loss rates were taken from the “Dutch” options. This includes several prescriptions: for O-stars the mass loss rates from \cite{2000A&A...362..295V, 2001V} are used; if the star enters the Wolf-Rayet stage, i.e. when the surface hydrogen mass fraction drops below 0.4, the mass loss rate switches to the scheme from \cite{2000A&A...360..227N}; if $T_\text{eff} < 10^4$ K, the empirical mass loss rate from \cite{1988A&AS...72..259D} is used. The mixing-length-theory \citep[MLT,][]{1958ZA.....46..108B} of convection describes the treatment of convection in our model (using the “Henyey” and “MLT++” options), where we applied an efficiency of $\alpha_\text{MLT}=1.67$ \citep{2018arXiv181004659A}. The Schwarzschild criterion defines the convective boundaries in our model and as such we did not need to implement semi-convective mixing. For convective boundary mixing, we included the exponential decaying diffusive model \citep{1996A&A...313..497F,2000A&A...360..952H} with $f_\text{ov} = 0.05$ for the top of convective cores and shells, and with $f_\text{ov} = 0.01$ for the bottom of convective shells (with $f_0 = f$ for both cases). We furthermore used the decline of the diffusion coefficient near the boundary \citep{2017MNRAS.465.2991J}. The value of $f_\text{ov} = 0.05$ for the top boundaries is larger than the majority of published large grids of stellar models (e.g.\ using $\alpha_\text{ov}=0.1$ in \citealt{2012A&A...537A.146E}, $\alpha_\text{ov}=0.335$ in \citealt{2011B}). The value of 0.05 is motivated by the study of \cite{2021MNRAS.503.4208S}, where values for $f_\text{ov}$ up to at least 0.05 for 20 M$_\odot$ and above (see caption of their Fig.\ 9 and references therein for the relation between $\alpha_\text{ov}$, $f_\text{ov}$ and entrainment) best reproduce the observed width of the main sequence in the spectroscopic HRD \citep{2014A&A...570L..13C}. For the bottom boundary, a CBM value of 1/5 the value of the top boundary is based on 3D hydrodynamic simulations \citep{2017MNRAS.471..279C, 2019MNRAS.484.4645C, 2022R} finding that CBM is slower at the bottom boundary due to it being stiffer and therefore harder to penetrate. \cite{2021MNRAS.503.4208S} also show that the amount of CBM should increase with initial mass since more massive stars are much more luminous ($L\sim M^3$ between 1 and 20 M$_\odot$) and thus our chosen value of 0.05 for 20 M$_\odot$ is consistent with the range of values inferred from asteroseismology for less massive stars \citep[$f_\text{ov} = 0.02-0.04$,][]{2020B}. As we describe below, the inclusion and the extent of CBM, guided by both 3D hydrodynamic simulations and observations (main-sequence width in HRD and asteroseismology) is a key aspect for the novelty of the results.

\subsection{The 3D hydrodynamic model setup}
The 1D simulation has been run from the pre-main-sequence phase until core collapse (see the structure evolution diagram in Fig.~\ref{fig:kip}, left). The hydrodynamic simulations have been started with initial conditions taken from the first neon-burning shell that develops in the 1D simulation (see Fig.~\ref{fig:kip}, right). We employ here the stellar hydrodynamical code \texttt{PROMPI} \citep{2007ApJ...667..448M}, successfully used over the years to simulate and study convection and CBM in advanced phases of massive stars \citep[e.g.][]{Arnett_2009,2017MNRAS.471..279C,2019MNRAS.484.4645C,2018M,2022R}. \texttt{PROMPI} has been also recently compared to other stellar hydrodynamical codes in \cite{2022A&A...659A.193A}, who showed \texttt{PROMPI} to be fully consistent with the other codes. \\
The radial variables density, pressure, temperature, entropy, mass and chemical composition have been remapped on a tri-dimensional grid adding relatively small perturbations ($\sim$$10^{-7}$) to density and temperature as seeds for convective instabilities. The grid we used is in spherical coordinates with a radial extension from $3.6$ to $8.5\times10^8$ cm and angular size of about $26^\circ$ in both angular dimensions. Since the radial size of the grid is roughly twice the size of the other dimensions, we used a resolution with double the cells in radius. Making use of the spherical coordinates, gravity is recomputed at each time step by integrating the mass inside each shell as function of the radius. This allows for contraction or expansion of the layers. A resolution of $256\times128^2$ grid points in $r,\theta$ and $\phi$, respectively, was initially used to explore the evolutionary timescale of the simulations, then for this study we analysed a more detailed set of simulations with resolution $512\times256^2$, and finally results have been validated with our most detailed simulations of $1024\times512^2$ and $2048\times1024^2$. \\
Convection has been fuelled with nuclear energy. Nuclear burning and nucleosynthesis are not always explicitly tracked in hydrodynamic simulations, because of the high computing cost: instead, fixed heating profiles are often used to drive convection. \cite{2022R} included a simple energy generation routine of 5 isotopes ($^4$He, $^{16}$O, $^{20}$Ne, $^{24}$Mg, $^{28}$Si) to reproduce neon burning. We have now extended this network to include 12 isotopes: n, p, $^{4}$He, $^{12}$C, $^{16}$O, $^{20}$Ne, $^{23}$Na, $^{24}$Mg, $^{28}$Si, $^{31}$P, $^{32}$S, $^{56}$Ni. We employed the most recent nuclear rates from the \texttt{JINA REACLIB} database \citep{Cyburt_2010}. While this extension does not have important effects on the energy release, which was already accurate with the 5-isotopes burning routine, it allows us to study the nucleosynthesis and transport of other species, paving the way towards an extended multi-D nucleosynthesis in stellar models. \\
Finally, following the same approach as \cite{2022R}, in this study we included  simulations with and without modified energy generation. In the nominal-luminosity simulations, the luminosity remains the same as in the 1D model. In this way, we make sure to validate our results ruling out the possibility that the boosting in luminosity introduces additional differences from the nominal case. When a boosting in luminosity was included, the nuclear rates for neon burning $^{20}$Ne$(\gamma,\alpha)^{16}$O and $^{20}$Ne$(\alpha,\gamma)^{24}$Mg have been multiplied by a boosting factor. Since these are the reactions that dominate the energy release, it does not make any difference that the other reactions have not been boosted. The boosting factors used here are 1 (nominal-luminosity), 5, 10 and 50. 

\begin{figure*}
\centering
\footnotesize
\includegraphics[trim={0cm 0 0cm 0cm},clip,width=0.75\textwidth]{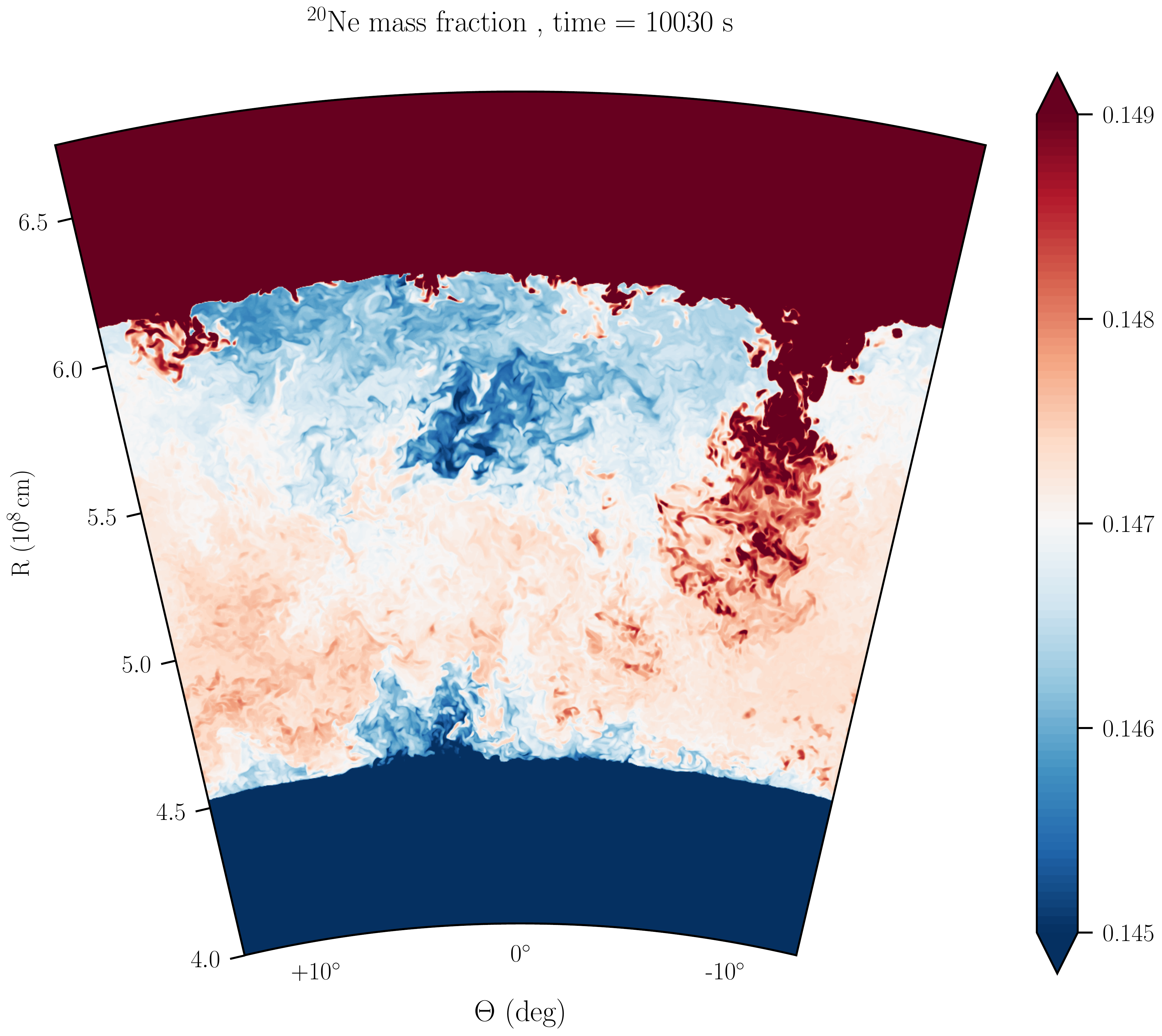}
\caption{Cross section of the neon mass fraction (values in colour scale) from the \texttt{r2048e1} simulation. The frame shows entrainment of some neon-rich material from the upper stable region into the convective zone. Videos of the evolution of the \texttt{r1024e1} simulation are available online as Supplementary material, showing fluid speed and mass fractions in colour scale.}\label{fig:sect}
\end{figure*}
\newpage
\subsection{Entrainment law computation}
\texttt{PROMPI} has a long history of modelling entrainment in stellar environments, having investigated a massive-star O-shell \citep{2007ApJ...667..448M}, C-shell \citep{2019MNRAS.484.4645C}, and Ne-shell \citep{2022R}. 
We can express the entrainment rate for a convective boundary, i.e.\ entrainment velocity $v_\text{e}$ divided by convective velocity $v_\text{rms}$, according to \citet{2007ApJ...667..448M}:
\begin{equation}
E = \dfrac{v_\text{e}}{v_\text{rms}} = A\ Ri_\text{B}^{-n}
\end{equation}
as a function of the “bulk Richardson number” $Ri_\text{B}$, that we can see as a measure of boundary resistance against convective fluid penetration, and defined as:
\begin{equation}
Ri_B=\dfrac{\ell\cdot\Delta b}{v_\text{rms}^2}\ ; \quad \Delta b=\int\limits_{r_\text{b}-\ell/2}^{r_\text{b}+\ell/2} N^2 dr
\end{equation}
with $\Delta b$ the buoyancy jump across the convective boundary, $N$ the Brunt–Väisälä frequency, $\ell$ the length scale of turbulent motions, and $r_\text{b}$ the convective boundary location.\\
For the computation of the quantities just defined, we refer to \cite{2019MNRAS.484.4645C} and \cite{2022R} since we used here their same approach and definitions. It is worth mentioning here that $r_\text{b}$ has been obtained using the neon abundance gradient, that $v_\text{e}$ is the time derivative of $r_\text{b}$, that $v_\text{rms}$ is equal to $\left(v_r^2+v_\theta^2+v_\phi^2\right)^{1/2}$ inside the convective zone, that $\ell$ has been set equal to 1/12 of the local pressure scale height to enclose the peak in $N^2$, and that all quantities have been averaged for the entire time the simulations spent in the entrainment regime.

\section{Results}\label{sec:3}
In this article, we present a set of 3D hydrodynamic simulations run continuously from the beginning to the end of a neon-shell burning phase. The level of realism of these new simulations is guaranteed by the updated 1D model for initial conditions, that includes a stronger CBM than usually implemented, the complex burning routine including 12 isotopes for energy generation, the presence of unaltered nominal-luminosity runs, spherical geometry, and some high-resolution runs. \\
In Table~\ref{tab:1}, we summarize the most important properties of the hydrodynamic simulations we have run for this study, each of them used for some part of the analysis, contributing to the final results we present here. The code name for each simulation summarises its radial resolution (\texttt{r}) and the boosting factor (\texttt{e}).

\begin{figure*}
\centering
\footnotesize
\begin{minipage}{0.5\textwidth}
\includegraphics[trim={0cm 0cm 0cm 0cm},clip,width=\textwidth]{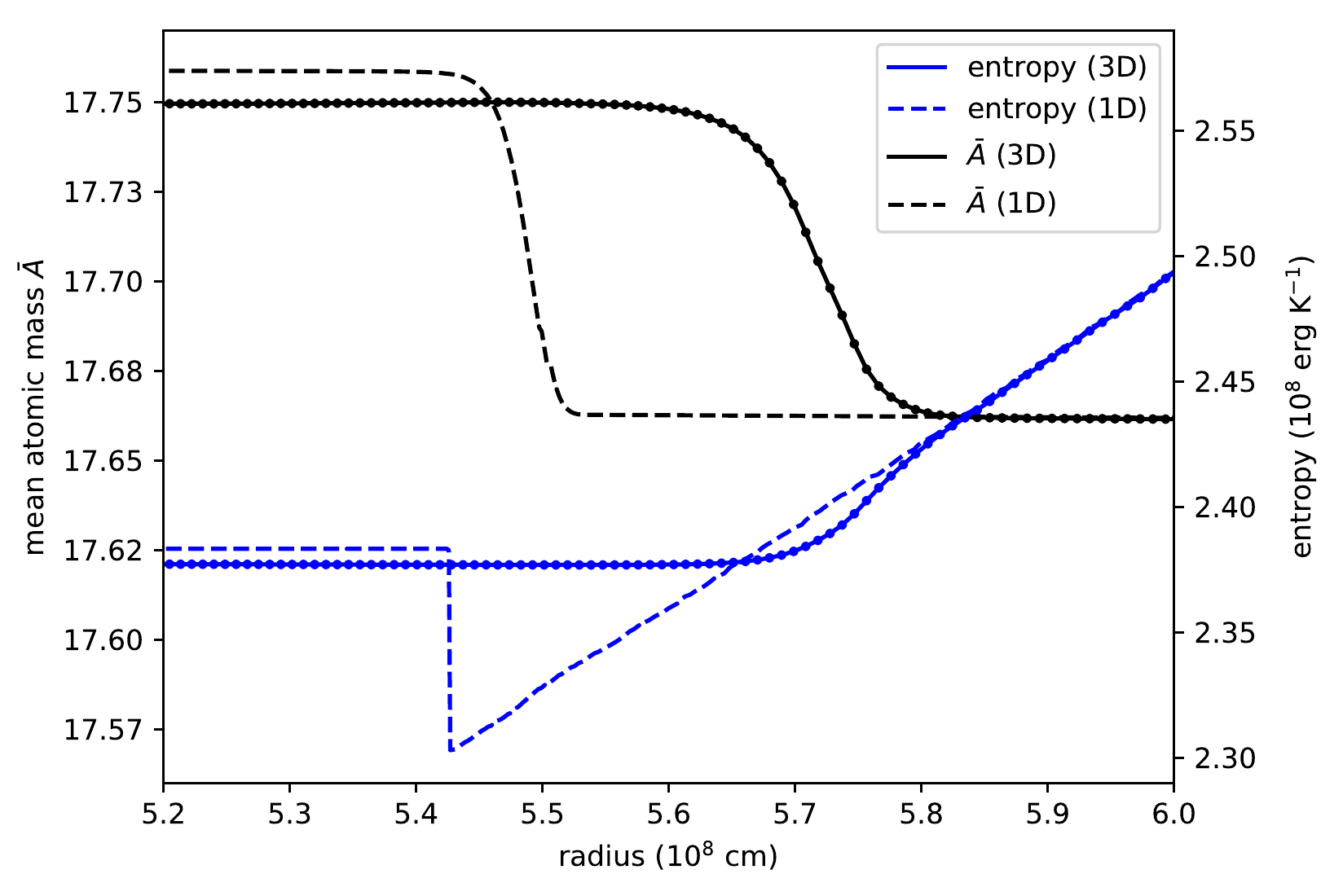}
\caption{Radial profiles of the mean atomic mass $\bar{A}$ (black) and entropy (blue) for the 1D input model (dashed lines) and at the end of a 3D \texttt{r512e1} test simulation (angularly-averaged quantities plotted as solid lines) run for 5 convective turnovers. The dots on the 3D curves indicate the location of the simulation cells.}\label{fig:abarentr}
\end{minipage}\hfill
\begin{minipage}{0.45\textwidth}
\centering
\footnotesize
\includegraphics[trim={0cm 0 0cm 0cm},clip,width=\textwidth]{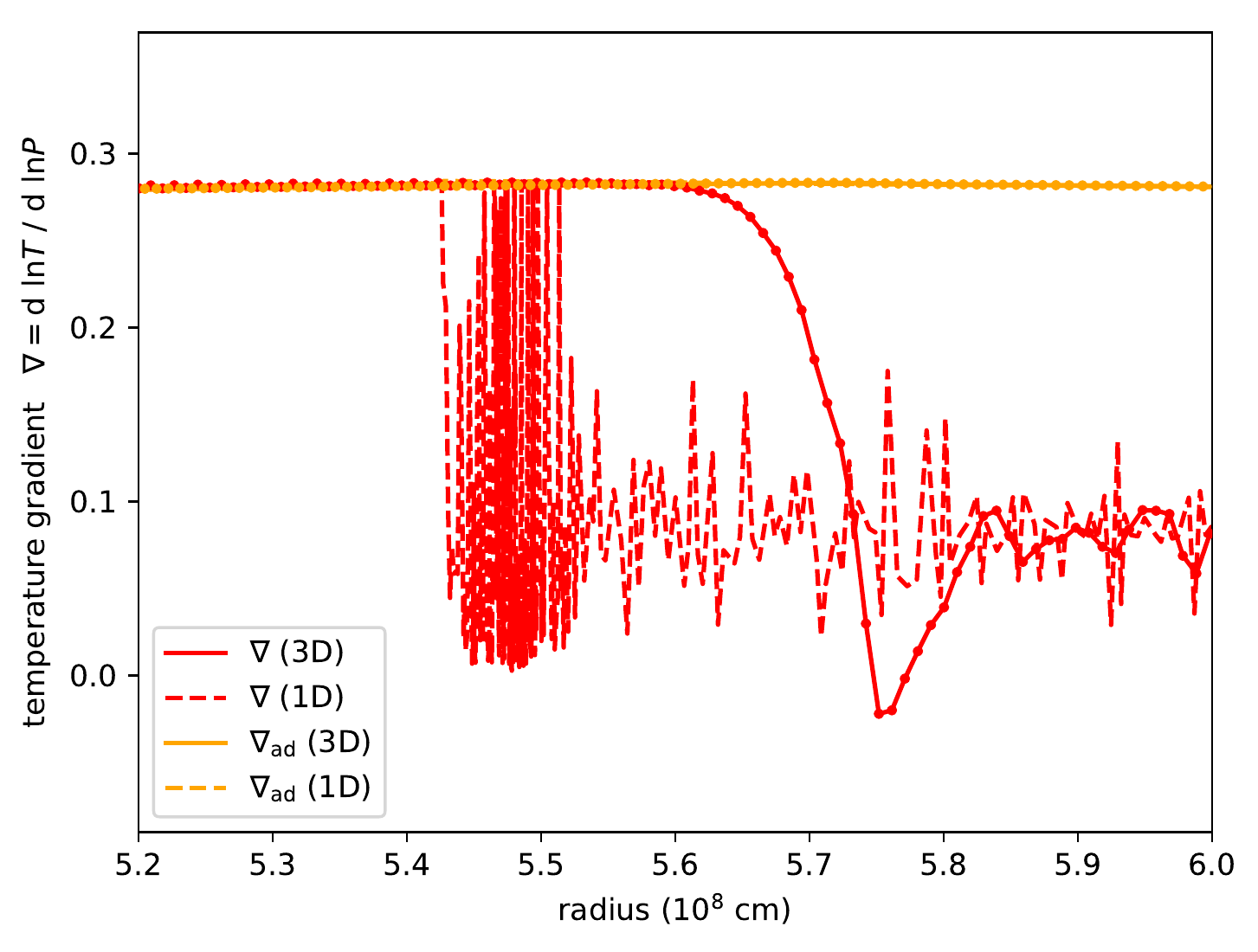}
\caption{Same as Fig.~\ref{fig:abarentr}, but radial profiles of the actual temperature gradient (red) and the adiabatic temperature gradient (yellow), defined as $\nabla=\text{d}\ln T/\text{d}\ln P$, for the 1D input model (dashed lines) and the 3D \texttt{r512e1} test simulation (angularly-averaged quantities plotted as solid lines).}\label{fig:grads}
\end{minipage}
\end{figure*}

\subsection{Analysis of the fluid dynamics}
\begin{figure*}
\centering
\footnotesize
\includegraphics[trim={0cm 0 0cm 0cm},clip,width=1.\textwidth]{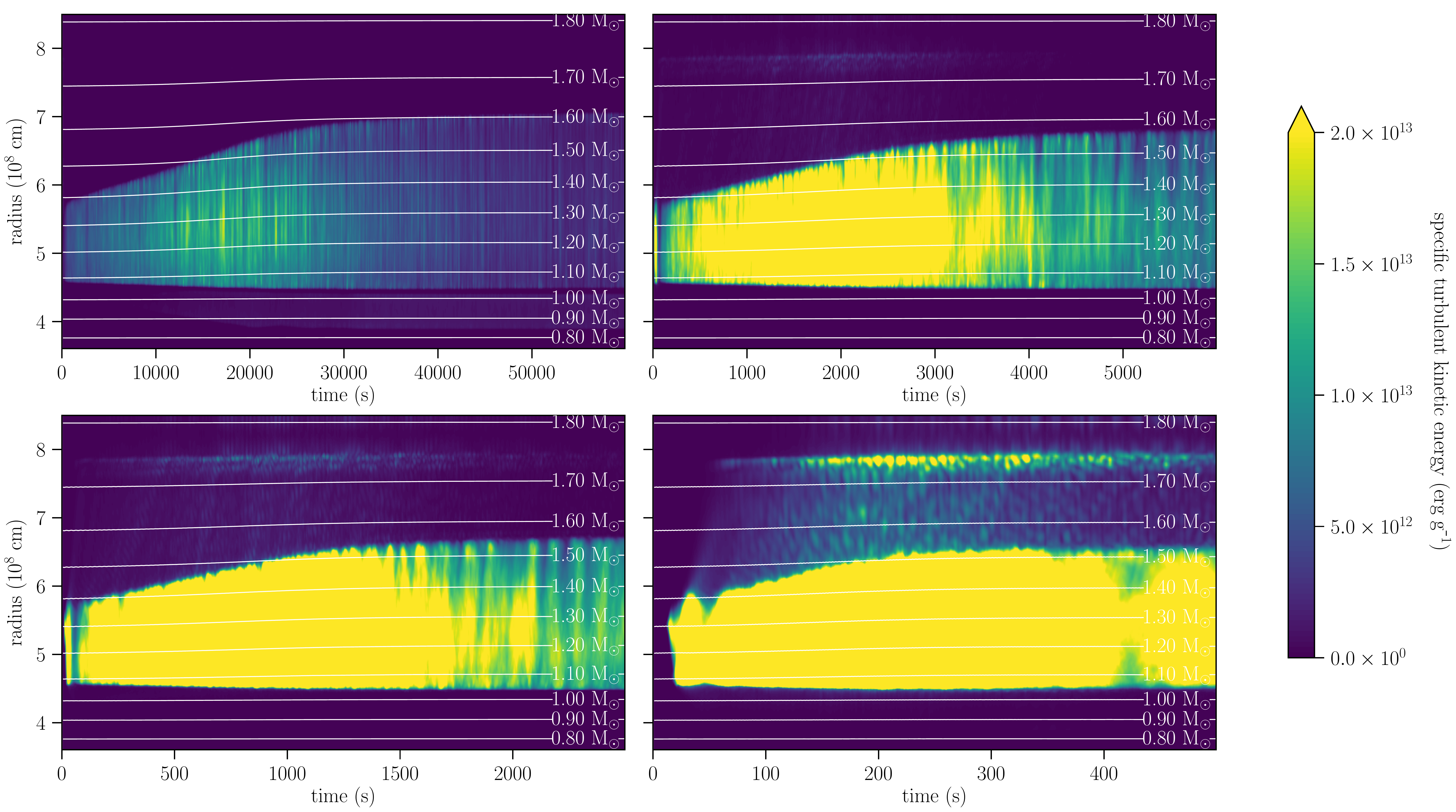}
\caption{Time evolution of the angularly-averaged specific turbulent kinetic energy (in colour scale) for four simulations with different luminosity boosting factors: from top left to bottom right, \texttt{r256e1}, \texttt{r256e5}, \texttt{r256e10}, \texttt{r256e50}. Overlaid in white, the isomass contours.}\label{fig:4pan}
\end{figure*}

To provide a visual representation of our simulations, we show in Fig.~\ref{fig:sect} a cross section of the neon mass fraction from our high-resolution nominal-luminosity \texttt{r2048e1} simulation. In addition to showing the fine detail that hydrodynamic simulations can reveal with modern computing resources, this image helps us understand the importance of CBM for stellar evolution. Indeed, it is possible to see the mixing of neon-rich and neon-poor material inside the convective zone, as well as the entrainment of neon-rich material from the upper stable region, due to shear mixing at the interface of the two layers. The entrainment of fresh fuel extends the nuclear burning timescale and therefore the convective shell lifetime, showing the importance of including CBM in all stellar models.\\
The effects of the mixing on the convective boundaries can also be seen in Fig.~\ref{fig:abarentr}, where we show the difference in the radial profiles of mean atomic mass and entropy between the 1D initial conditions and a 3D \texttt{r512e1} test simulation. This 3D \texttt{r512e1} test simulation was run for 5 convective turnovers to highlight similarities and differences between the 1D and 3D profiles near the start of the simulations (the \texttt{r512e1} simulation listed in Table\,\ref{tab:1} and discussed in the rest of the paper was restarted from the \texttt{r256e1} simulation after 16,000 seconds to make the most of our computing budget). We added in Fig.~\ref{fig:abarentr} the grid points on the 3D simulation profiles, to indicate the resolution of the simulation across the convective boundary. The reason why the values for atomic mass and entropy in the convective plateaus at $r<5.4\times 10^8$ cm do not match perfectly ($\sim10^{-4}$ absolute difference) is the remapping and the small mixing that takes place during the initial transient. Importantly, what we can see from this plot is that, as the boundary moves outwards due to CBM, both chemical composition and entropy are consequently mixed. This point is underlined also by Fig.~\ref{fig:grads}, which shows the temperature gradients for the same models. Since both 1D and 3D stellar models deviate from the adiabatic temperature gradient outside the convective region (although the 1D curve is affected by numerical noise), it is clear that CBM in the 3D simulation contributes to altering the temperature gradient in the ``overshooting'' region, which becomes adiabatic due to the fast entropy mixing. These plots show that in the 1D \texttt{MESA} code the composition mixing is accurately modelled, since the shape of $\bar{A}$ is similar to the 3D, but the entropy profile is not compatible, so it will be necessary in the future to also add the mixing of entropy in the 1D models.
\begin{figure*}
\centering
\footnotesize
\includegraphics[trim={0cm 0 0cm 0cm},clip,width=1\textwidth]{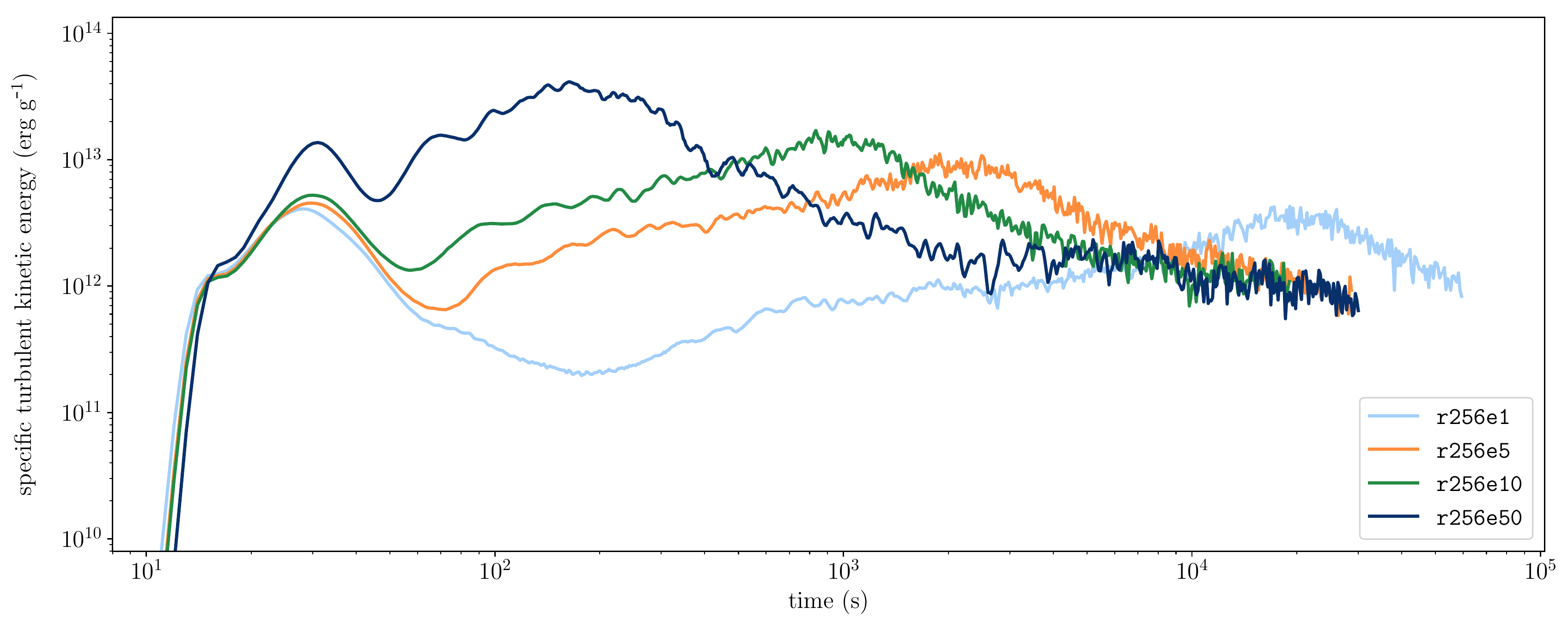}
\caption{Time evolution of the specific turbulent kinetic energy $\dfrac{1}{2}\left(\overline{v_r^2}-\overline{v_r}^2+\overline{v_\theta^2}-\overline{v_\theta}^2+\overline{v_\phi^2}-\overline{v_\phi}^2\right)$ for the four simulations with different luminosity boosting \texttt{r256e1}, \texttt{r256e5}, \texttt{r256e10}, \texttt{r256e50}.}\label{fig:tke}
\end{figure*}
\\Another way of studying the evolution of the different simulations is to analyse how the kinetic energy of the fluid builds up and changes with time. In Fig.~\ref{fig:tke} we show the time evolution of the specific turbulent kinetic energy, defined as $\dfrac{1}{2}\left(\overline{v_r^2}-\overline{v_r}^2+\overline{v_\theta^2}-\overline{v_\theta}^2+\overline{v_\phi^2}-\overline{v_\phi}^2\right)$ where the single quantities have been averaged in $r,\theta$ and $\phi$ inside the convective zone. The figure shows the most complete set of simulations we have run, which consists of \texttt{r256e1}, \texttt{r256e5}, \texttt{r256e10}, \texttt{r256e50}. These simulations have the same resolution but different luminosity boosting, and this explains the different evolution of the tracks in the figure. After an initial transient, during which convection develops in the 3D grid and the kinetic energy builds up, all simulations evolve in a similar way, with a more or less gradual increase in kinetic energy during the nuclear burning phase, and a slow decrease after neon is consumed and nuclear burning does not sustain the kinetic energy any longer. The differences in magnitude and timescale can be all traced back to the boosting in luminosity, which strongly affects both the amount of kinetic energy produced during the simulation, and how rapidly neon is consumed, therefore the timescale for nuclear burning. The peak in kinetic energy is approximately at 20000, 2200, 960 and 180 seconds for \texttt{r256e1}, \texttt{r256e5}, \texttt{r256e10} and \texttt{r256e50} respectively.\\
It is also interesting to notice that towards the end all simulations seem to converge to the same constant value of turbulent kinetic energy, regardless of the luminosity boosting. This comes from the fact that after neon is exhausted the reactions $^{20}$Ne$(\gamma,\alpha)^{16}$O and $^{20}$Ne$(\alpha,\gamma)^{24}$Mg stop occurring, so the only nuclear reaction that can proceed is the secondary $^{24}$Mg$(\alpha,\gamma)^{28}$Si, which was not boosted and has the same rate for all simulations. This is confirmed by a slight decrease in magnesium and increase in silicon abundances during this late phase. However, this reaction is not energetic enough, therefore turbulence decays and the shell growth halts.
\\The time evolution of the turbulent kinetic energy is also shown in Fig.~\ref{fig:4pan} for the same set of simulations, but as a colour map with angularly-averaged values in colour scale. First, it is interesting to note the strong effect that the luminosity boosting has on the evolutionary timescale. Apart from this, the shell is evolving in the same way in all simulations, with the first period being dominated by nuclear burning and entrainment, and demonstrating a linear growth of the shell. An additional weak burning front is visible around $r\sim8\times10^8$ cm, produced by the impact of gravity waves on a carbon shell above the neon one, but its energy is so small compared to the neon burning shell (three orders of magnitude lower), that it has no impact on convection and entrainment. Once fuel is exhausted in the neon-burning shell (neon abundance is 6\% at the peak of kinetic energy) convection slowly diminishes, as evident from the drop in kinetic energy, and therefore entrainment also stops, halting the shell growth. This result has important implications for stellar evolution. In contrast to previously suppositions \cite[e.g.][]{2019MNRAS.484.4645C,2021A&A...653A..55H,2022R}, CBM does not lead to the convective engulfment of the entire star, but our simulations show that the shell naturally stops growing when its fuel is exhausted. Indeed, making use of network calculations with a one-zone model to exclude convective mixing, we estimated the nuclear burning timescale ($X_\text{Ne}/\dot{X}_\text{Ne}$) to be around 4000 seconds in this environment, which is much shorter than the timescale for mass entrainment ($M_\text{e}/\dot{M}_\text{e}$), here around 30000 seconds, therefore entrainment cannot sustain convection on its own. This finding puts a limit on the size of convective zones, which can have a strong effect on the final structure of massive stars.
\\There are other important points that can be drawn from Fig.~\ref{fig:4pan}. Comparing the nominal-luminosity run (first panel) to the corresponding evolution in the 1D model (Fig.~\ref{fig:kip}, right), it is evident that the convective shell in the 3D model evolves approximately $\sim$5 times faster than in its 1D equivalent. To understand this difference, in Fig.~\ref{fig:4pan} we overlay isomass contours in white that can be used to track expansion. The contours show some expansion in the bulk of the convective zone, but no expansion is allowed closer to the upper and lower domains, because mass flow is not allowed through those closed boundaries. If compared again to Fig.~\ref{fig:kip} (right), it is clear that an expansion is present in the 1D layers that are not limited by domain boundaries, but which are free to contract or expand. It is this difference in convective zone size between 1D and 3D that explains the difference in burning timescales. Indeed, the neon-burning energy release is strongly dependent on the temperature, sensitive to a power of $\sim T^{50}$ \citep{2002W} due to the temperature dependence of nuclear reaction rates and the $\alpha$-particle mass fraction. Furthermore, the temperature of a gas is dependent on its volume, according to the equation of state. 
It is possible to show with a simple calculation that the volume difference between the final states of our convective zone in 1D versus 3D 
can explain the shorter timescale of the 3D simulation. If we assume for simplicity that the two states are separated by an adiabatic expansion (it is reasonable to assume no heat exchange with the surroundings), then:
\begin{equation}
\dfrac{T_\text{3D}}{T_\text{1D}} =\left(\dfrac{V_\text{1D}}{V_\text{3D}} \right)^{\gamma-1} =\left(\dfrac{R^3_\text{1D}-r^3_\text{1D}}{R^3_\text{3D}-r^3_\text{3D}} \right)^{\gamma-1}
\end{equation}
where $r$ and $R$ are the inner and outer radii of the shells, respectively.
Comparing the 1D and 3D states at the end of the nominal-luminosity neon burning (taken when the neon abundance in the bulk of the convective zone $X_\text{Ne}\sim0.015$), we have (in units of $10^8$ cm):
\begin{equation*}
R_\text{1D}=7.34,\quad r_\text{1D}=4.63,\quad  R_\text{3D}=6.93,\quad r_\text{3D}=4.44
\end{equation*}
and $\gamma=1.55$, which gives $T_\text{3D} /T_\text{1D}=1.11$. 
At the same point in the actual simulations, we find $T_\text{3D}=1.88$\,GK and $T_\text{1D}= 1.78$\,GK at the temperature peak (located near the bottom of the convective shell) corresponding to a ratio of $1.06$. Thus, we can conclude that the limited expansion in 3D due to the closed boundary conditions can account for the higher temperature found in the 3D simulations compared to the 1D stellar model. \\
Since the nuclear energy generation rate depends on a power of the temperature $\dot{\epsilon}\sim T^\alpha$, the difference in nuclear burning timescale $\tau\sim 1/\dot{\epsilon}$ between 1D and 3D can be approximated as:
\begin{equation}
\dfrac{\tau_\text{3D}}{\tau_\text{1D}} \sim\left(\dfrac{T_\text{1D}}{T_\text{3D}} \right)^\alpha
\end{equation}
Using $\alpha=50$ from \cite{2002W} results in a 3D timescale 15 times shorter than for the 1D, which is much faster than what we see in our simulations. Realistically, the energy generation rate is also dependent on the neon abundance, which decreases with time, therefore $\alpha<50$ is expected, although the value changes in time. While both the 1D and 3D simulations are complex, this simple estimate of the effect on the temperature of the different expansion shows that this can explain the difference in the timescales of neon burning between the 3D and 1D simulations.

\subsection{Spectral analysis and turbulence theory}
\begin{figure}
\centering
\footnotesize
\includegraphics[trim={0cm 0 0cm 0cm},clip,width=0.48\textwidth]{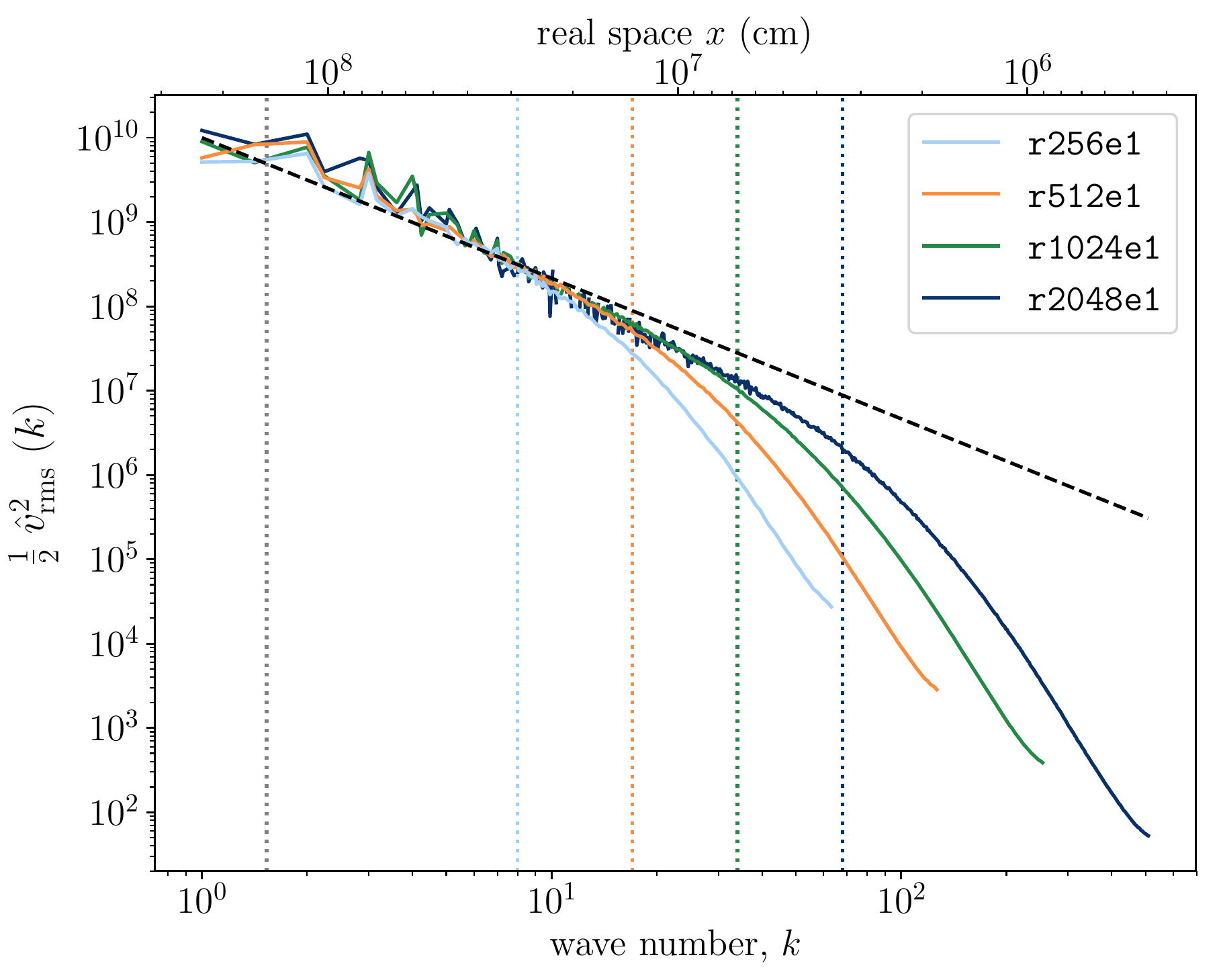}
\caption{Spectra of the specific kinetic energy as function of both the wave number $k$ and the real space $x$, taken on a surface at fixed radius, for the simulations with different resolution \texttt{r256e1}, \texttt{r512e1}, \texttt{r1024e1}, \texttt{r2048e1}. The spectra were averaged over 200\,s, except for \texttt{r2048e1} that was averaged over 10\,s. The dashed black line is the Kolmogorov scaling $k^{-5/3}$; the vertical dotted line at $k\sim 2$ is the size of the convective region; the vertical dotted lines at $k\sim 8$ - 70 correspond to the size of 15 cells for each simulation.
}\label{fig:spec1}
\end{figure}

\begin{figure}
\centering
\footnotesize
\includegraphics[trim={0cm 0 0cm 0cm},clip,width=0.48\textwidth]{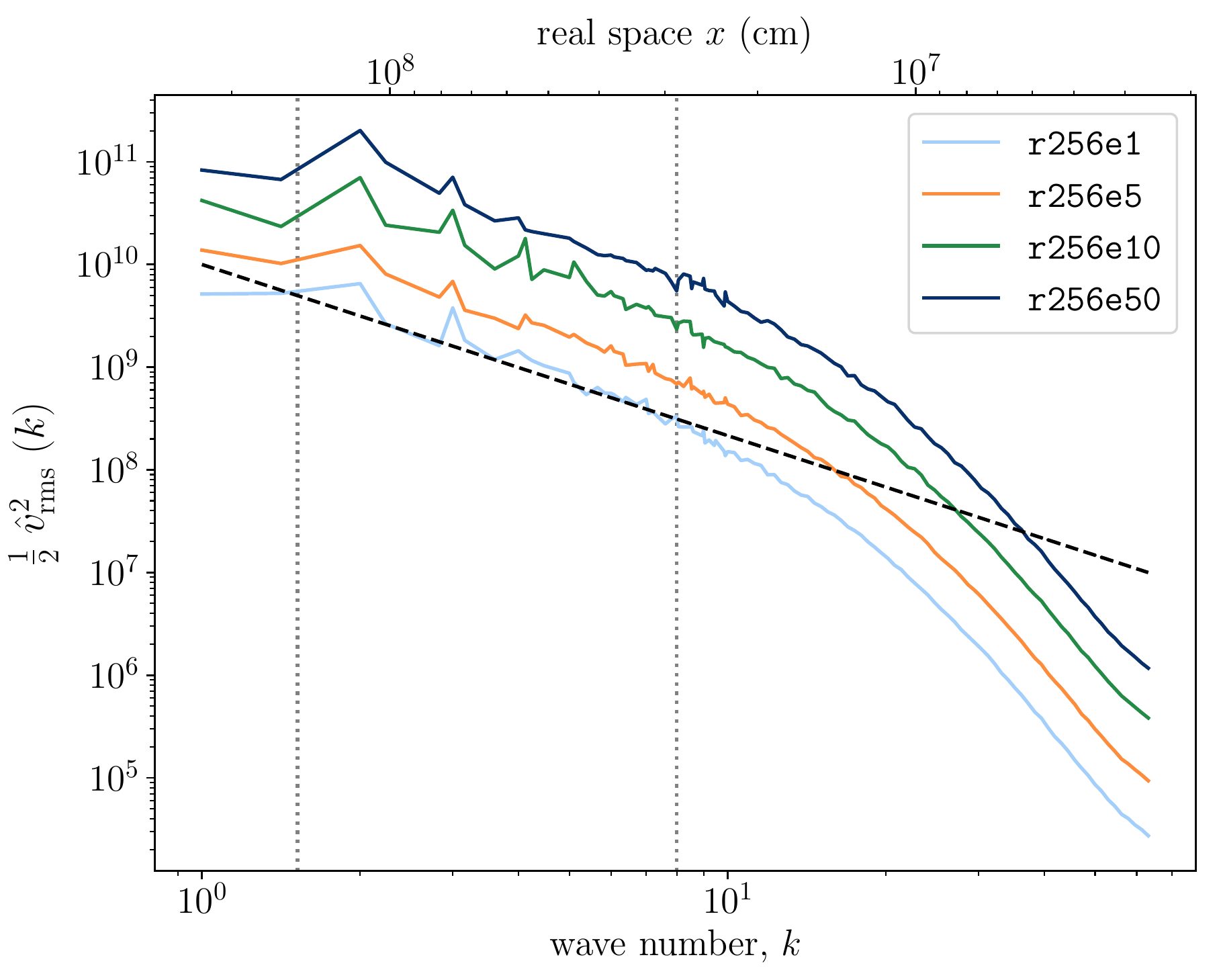}
\caption{Same as Fig~\ref{fig:spec1}, but for simulations with same resolution and different boosting factors: \texttt{r256e1}, \texttt{r256e5}, \texttt{r256e10}, \texttt{r256e50}. All spectra were averaged over one convective turnover. The two vertical lines identify the inertial range, as in Fig~\ref{fig:spec1}.}\label{fig:spec2}
\end{figure}

The well-established ``Kolmogorov theory'' \citep{1941DoSSR..30..301K} states that, for a quasi-steady isotropic regime of convection, the rate of energy dissipation is independent of the scale and of the type of dissipative process, and the kinetic energy is expected to behave according to:
\begin{equation}
E_\text{K}\sim v^2_\text{rms}\sim k^{-5/3}
\end{equation}
where $k$ is the wave number associated with the fluid scale. The fluid is expected to follow this scaling throughout the so-called ``inertial range'', while it deviates from it at the smallest (dissipation) scales due to dissipative effects, and at the largest (integral) scales because it stops being isotropic. These premises allowed hydrodynamic codes to employ the implicit large eddy simulation (ILES) scheme, that replaces the explicit physical viscosity with  implicit numerical viscosity due to the finite grid resolution. An important advantage of ILES is that it overcomes the necessity of resolving the flow at the viscosity scale, which would be impossible for stellar simulations. \\
It is thus possible to determine the quality of our simulations in the inertial range by comparing the power spectrum of the turbulent velocity in the simulations to the  scaling expected from Kolmogorov theory. Since we are employing spherical geometry, the default method would be to compute the spherical harmonic decomposition of $v_\text{rms}$. However, our simulations cover only a small section of a spherical surface (0.2 steradians, or 2\% of the total spherical surface), therefore it is not trivial to do the decomposition, which would also be unable to show the lowest-order modes due to the limited solid angle covered. One way of doing that would be to repeat the pattern periodically to fill a full spherical surface, as done in \cite{2021A&A...653A..55H}, but this would introduce artifacts coming from the spherical mapping and the risk of underestimating the low-order modes. For these reasons, we prefer here to compute the spectrum using a 2D Fourier analysis, following a similar approach to e.g.\ \cite{2019MNRAS.484.4645C}, \cite{2022A&A...659A.193A}. Selecting a radius, $r=5\times10^8$ cm, in the middle of the convective region, a 2D Fourier transform of the velocity magnitude as a function of the angular coordinates $\theta,\phi$ was computed:
\begin{equation}
\hat{v}_\text{rms}(k_\theta,k_\phi)=\dfrac{1}{N_\theta N_\phi}\sum_{n_\theta=0}^{N_\theta-1}\sum_{n_\phi=0}^{N_\phi-1}v_\text{rms}(\theta,\phi)e^{-i2\pi\left(\dfrac{k_\theta n_\theta}{N_\theta}+\dfrac{k_\phi n_\phi}{N_\phi}\right)}
\end{equation}
where $N_\theta,N_\phi$ are the numerical resolution, $n_\theta,n_\phi$ the cell numbers, and $k_\theta,k_\phi$ the wave numbers, that span the range:
\begin{equation}
   \begin{aligned}
    &    k_\theta = \begin{cases}
        i, & \text{if\quad} 0 \le i < N_\theta/2 \\
        i-N_\theta, & \text{if\quad} N_\theta/2 \le i < N_\theta
    \end{cases}\\
    &    k_\phi = \begin{cases}
        j, & \text{if\quad} 0 \le j < N_\phi/2 \\
        j-N_\phi, & \text{if\quad} N_\phi/2 \le j < N_\phi
    \end{cases}
\end{aligned} 
\end{equation}
Finally, we plot in Figs.~\ref{fig:spec1} and \ref{fig:spec2} the power spectrum $\dfrac{1}{2}|\hat{v}_\text{rms}|^{2}$, that can be interpreted as the specific kinetic energy, as a function of the wave number $k=\sqrt{k_x^2+k_y^2}$. Since $k_\theta\in\left[-N_\theta/2,N_\theta/2\right]$ and likewise for $k_\phi$, and the norm $k$ draws a circle in the $(k_\theta,k_\phi)$ space, we limit the plot to $k\in[0,min\{N_\theta/2,N_\phi/2\}]$ to avoid the circle going beyond the domain and losing a fraction of the signal, resulting in a drop of the power spectrum. Spectra have been averaged over at least one convective turnover for all simulations, except \texttt{r2048e1} that has been run for a very short timescale given its very high resolution, so it has been averaged for the last 10 s. \\
Figure~\ref{fig:spec1} shows the spectra for simulations with boosting factor equal to 1 (nominal luminosity) but different resolutions. In all simulations, the bulk of the spectrum follows the expected Kolmogorov scaling, which is a good confirmation of the presence of a large inertial range in our simulations. Also, as expected, the spectrum starts deviating from the $k^{-5/3}$ scaling both at the largest scales (around the vertical $k\sim 2$ line) and at the smallest scales, because of the numerical dissipation near the grid scale. For this reason, as resolution increases in Fig.~\ref{fig:spec1} the inertial range extends towards larger wave numbers, because dissipation takes place on smaller and smaller spatial scales. The point where the spectrum slope starts deviating strongly from the $k^{-5/3}$ scaling corresponds approximately to 15 cells for all simulations, as we indicate in the plot with vertical dotted lines around $k\sim 8$ - 70.\\
In Fig.~\ref{fig:spec2} we present instead the spectra for a set of simulations with same resolution but different boosting factors. As expected, the specific kinetic energy of the spectra increases with the boosting factor, but the extent of the inertial range does not change, since the resolution is the same. This is a confirmation of the fact that introducing a boosting factor does not affect the general properties of the turbulent flow (apart from the magnitude of the kinetic energy).  These findings are perfectly in line with previous simulations \citep[e.g.][]{2019MNRAS.484.4645C}, and especially with results from the code comparison study of \cite{2022A&A...659A.193A}.


\subsection{Entrainment analysis and parametrization}
\begin{figure}
\centering
\footnotesize
\includegraphics[trim={0cm 0 0cm 0cm},clip,width=0.48\textwidth]{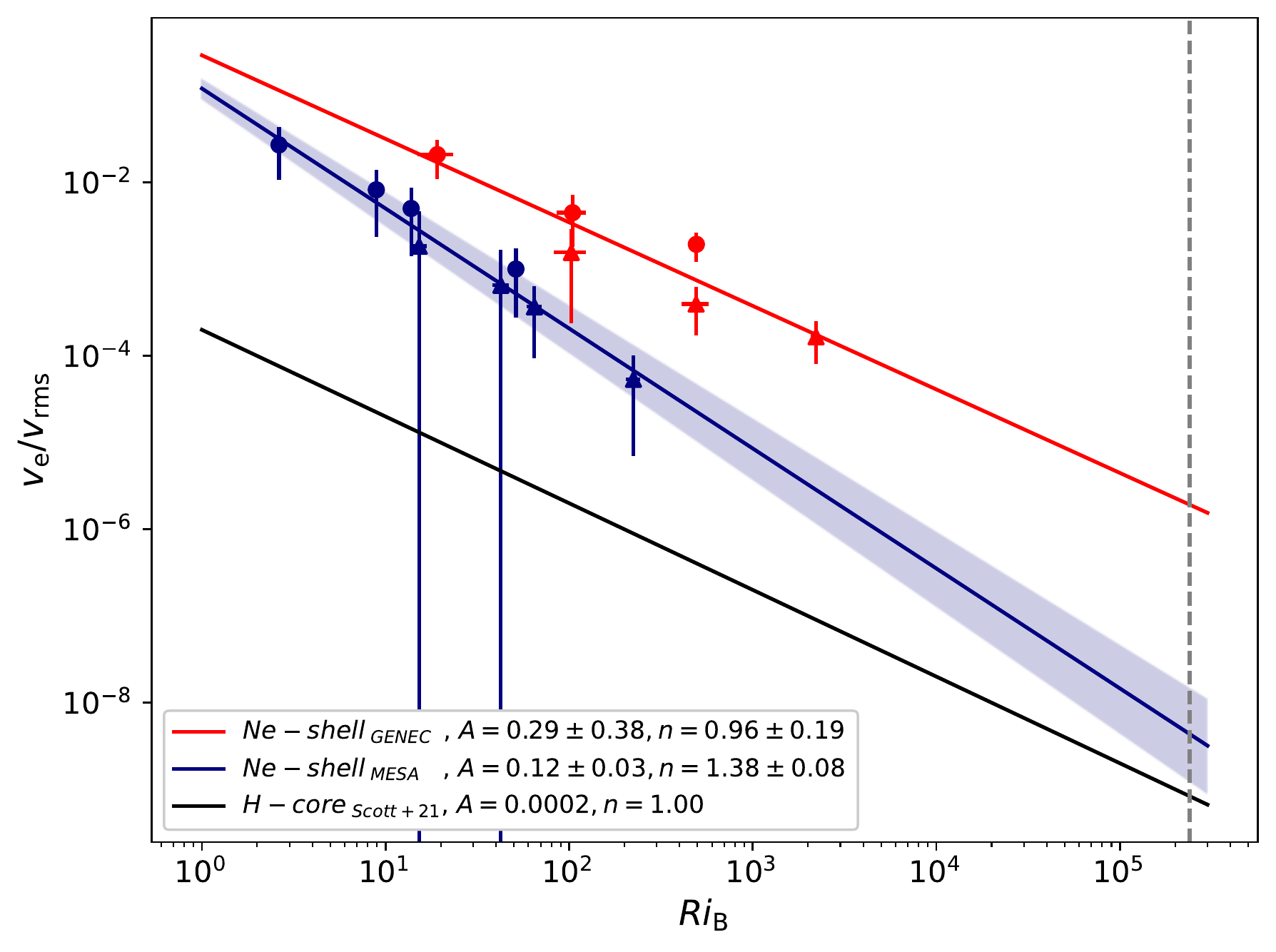}
\caption{Entrainment rate versus bulk Richardson number. Data from stellar simulations and respective linear regressions: “\texttt{MESA}” Ne-shell from this study (blue), “\texttt{GENEC}” Ne-shell from \protect\cite{2022R} (red), and H-core from 1D \protect\cite{2021MNRAS.503.4208S} (black). Triangles are lower convective boundaries, circles are upper boundaries. The dashed vertical line indicates the bulk Richardson number in the convective H-core. Error bars are standard deviations. In the legend, parameter estimates for the entrainment law $v_\text{e}/v_\text{rms} = A\ Ri_\text{B}^{-n}$.}\label{fig:laura}
\end{figure}

\begin{figure}
\centering
\footnotesize
\includegraphics[trim={0cm 0 0cm 0cm},clip,width=0.48\textwidth]{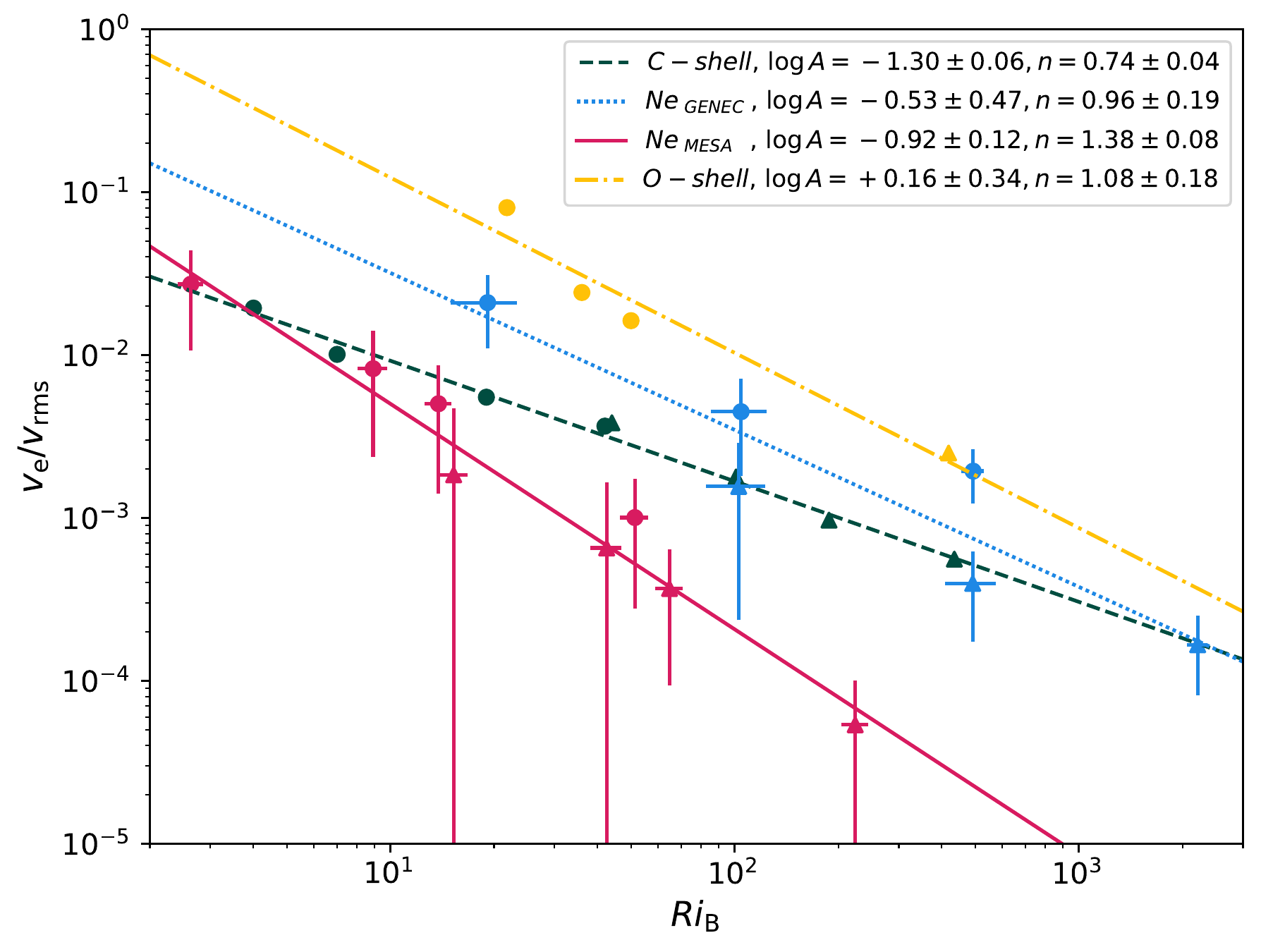}
\caption{Same as Fig.~\ref{fig:laura}, but comparison between \texttt{PROMPI} simulations: “\texttt{MESA}” Ne-shell from this study (red, solid), “\texttt{GENEC}” Ne-shell from \protect\cite{2022R} (blue, dotted), C-shell from \protect\cite{2019MNRAS.484.4645C} (green, dashed), O-shell from \protect\cite{2007ApJ...667..448M} (in yellow, dot-dashed). In the legend, parameter estimates for the entrainment law $v_\text{e}/v_\text{rms} = A\ Ri_\text{B}^{-n}$.}
\label{fig:entr}
\end{figure}

\begin{table}
\centering
\footnotesize
\caption{Properties of the 3D hydrodynamic simulations used to study entrainment in this work: model name; root-mean-square convective velocity $v_\text{rms}$ (cm s$^{-1}$); upper entrainment rate $v_\text{e}^\text{up}/v_\text{rms}$; lower entrainment rate $v_\text{e}^\text{low}/v_\text{rms}$; upper bulk Richardson number $Ri_\text{B}^\text{up}$; lower bulk Richardson number $Ri_\text{B}^\text{low}$.}\label{tab:2}
\setlength\tabcolsep{5pt}
\begin{tabular}{lccccc}
\hline
\hline
\texttt{name}&$v_\text{rms}$&$v_\text{e}^\text{up}/v_\text{rms}$&$v_\text{e}^\text{low}/v_\text{rms}$&$Ri_\text{B}^\text{up}$&$Ri_\text{B}^\text{low}$\\
\hline
\texttt{r512e1}&3.83 $\times$ 10$^6$&1.01 $\times$ 10$^{-3}$&5.38 $\times$ 10$^{-5}$&51.3&224\\
\texttt{r512e5}&6.65 $\times$ 10$^6$&5.03 $\times$ 10$^{-3}$&3.69 $\times$ 10$^{-4}$&13.8&64.7\\
\texttt{r512e10}&8.28 $\times$ 10$^6$&8.25 $\times$ 10$^{-3}$&6.54 $\times$ 10$^{-4}$&8.91&42.5\\
\texttt{r512e50}&1.34 $\times$ 10$^7$&2.72 $\times$ 10$^{-2}$&1.84 $\times$ 10$^{-3}$&2.63&15.3\\
\hline
\end{tabular}
\end{table}
As mentioned above, entrainment of fresh fuel into the convective zone is one of the most important effects on stellar convection, profoundly affecting the stellar structure and its evolution. We remind that the new simulations we are presenting in this work have been started from initial conditions taken from a 1D model using strong CBM at all convective boundaries, including the late-phase shells. This last point is particularly important in order to understand how entrainment differs between 1D and 3D stellar models, since many studies underline strong discrepancies between the two \citep{2013ARep...57..380S,2021Higl,2021A&A...653A..55H,2021MNRAS.503.4208S,2022R}. In particular, entrainment in 3D is always found to be much stronger than in 1D models, which often include little or no CBM at all. Starting from a 1D model with strong CBM, we can determine if the corresponding entrainment in 3D is larger as usual, or if a convergence between 1D and multi-D stellar models can be achieved.
\\In Fig.~\ref{fig:laura}, we present the entrainment rates estimated from the data presented in this work (blue), alongside one of the previous \texttt{PROMPI} simulations of a neon shell from \cite{2022R} (red), and the 1D study of a convective hydrogen core in a 15 M$_\odot$ star from \cite{2021MNRAS.503.4208S} (black). To study entrainment, we used the data coming from the \texttt{r512e1}, \texttt{r512e5}, \texttt{r512e10}, \texttt{r512e50} simulations, which have been run for many convective turnovers with a high resolution. For each simulation we computed the entrainment rate and bulk Richardson number for both the upper and lower convective boundaries, averaged through the entire entrainment phase; we list results in Table~\ref{tab:2}. As expected, a larger boosting factor in the simulations results in higher convective and entrainment velocities, and smaller $Ri_\text{B}$ due to the larger penetrability. In addition, the upper boundary has always larger entrainemnt rate and smaller $Ri_\text{B}$ than the lower one. Error bars in the figure are standard deviations of the values at each time-step in our simulations, and since both their computation and the fitting has been done in real space, in some cases the log scale of the plot shows the bars going towards zero.
\\It is evident from Fig.~\ref{fig:laura} that the value of $Ri_\text{B}$ for convection in late phases (data points) is several orders of magnitude smaller than during the main sequence (vertical line): the lack of entrainment data and the consequent need for extrapolation is the main reason for the current disagreement on CBM between different stellar phases. From the comparison of the different trends in Fig.~\ref{fig:laura}, several interesting conclusions may be drawn. The entrainment rates measured from our simulations are lower then all the previous multi-D studies done with \texttt{PROMPI}, as we show in Fig.~\ref{fig:entr} where we compare entrainement in all \texttt{PROMPI} simulations so far. Although our new rates are not as small as the ones predicted from studying the H-core in 1D (black line in Fig.~\ref{fig:laura}), the larger steepness and lower dispersion of the new results imply much less entrainment than the previous studies, especially at larger $Ri_\text{B}$, reaching a surprisingly good agreement with predictions for the convective core in 1D models (dashed vertical line). This finding is an important step towards convergence of results between 1D and 3D stellar models. \\
To better understand the reasons for this convergence, it is important to underline the differences between the previous and the new simulations. The two sets of hydrodynamic simulations we show in Fig.~\ref{fig:laura} are both of a Ne-burning shell, with a similar burning network and energy release, but with initial conditions taken from two different 1D models: one \citep[red]{2022R} from a \texttt{GENEC} model with no CBM for this phase, and the other (this study, blue) from a \texttt{MESA} model with strong CBM. The stellar mass is also different (15 M$_\odot$ for \citealt{2022R} and 20 M$_\odot$ for this study), but previous \texttt{PROMPI} simulations show that stellar mass does not seem to affect the entrainment law parameters \citep{2019MNRAS.484.4645C,2022R}. Furthermore, the present set of simulations has been run for much longer than before, covering the entire shell evolution until fuel exhaustion. What we can conclude from this is that a hydrodynamic simulation started from a 1D model already including CBM produces significantly less entrainment than simulations from a 1D model with no CBM. This is a clear sign of convergence in the old problem of comparing CBM between 1D and 3D models. Moreover, it is a significant confirmation of the potential of this novel approach towards developing 3D stellar evolutionary simulations.

\subsection{Nucleosynthesis and evolution of the chemical composition}
\begin{figure}
\centering
\footnotesize
\includegraphics[trim={0cm 0 0cm 0cm},clip,width=0.48\textwidth]{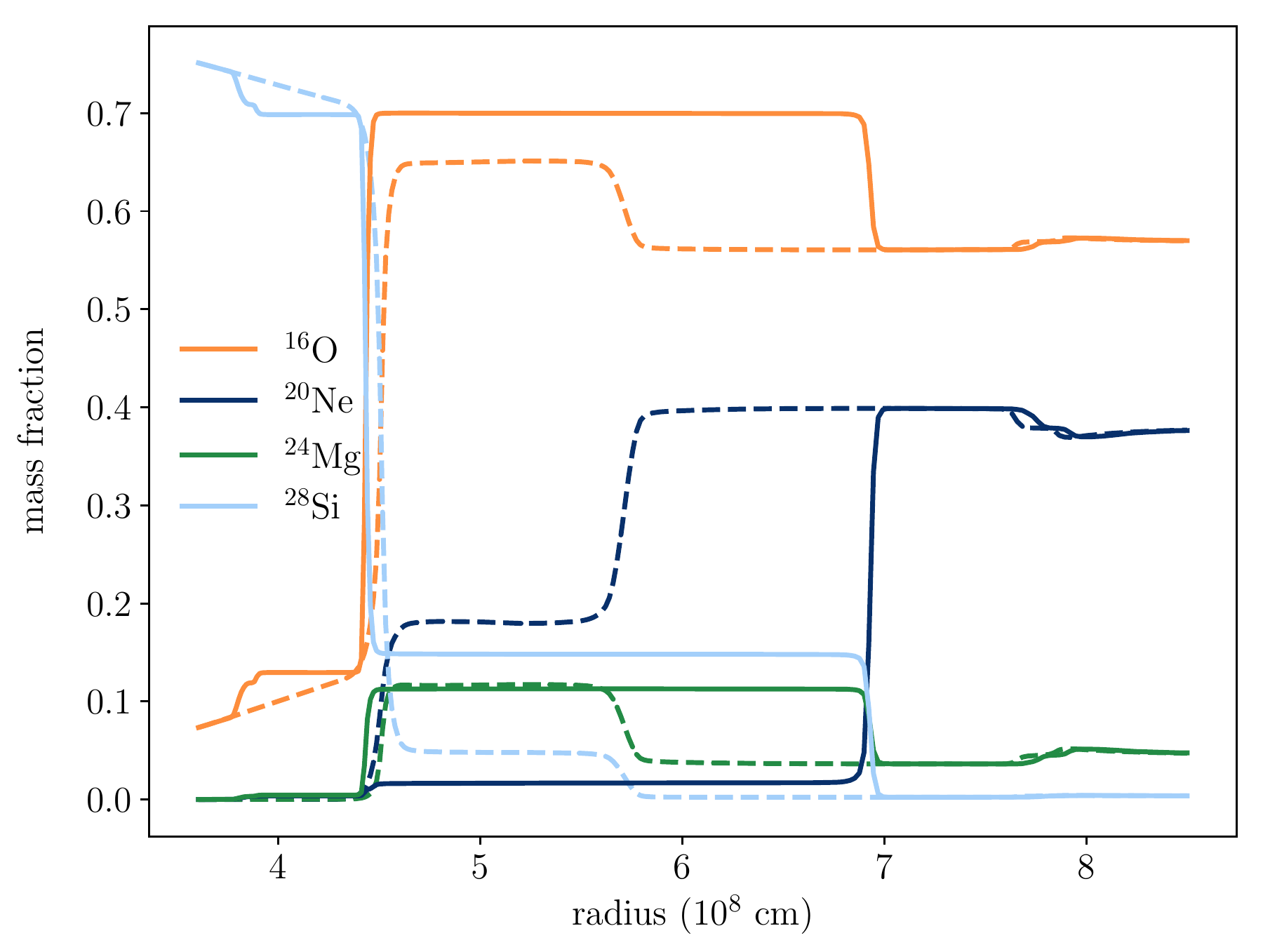}
\caption{Angularly-averaged mass fractions of $^{16}$O, $^{20}$Ne, $^{24}$Mg, $^{28}$Si as function of the stellar radius. Dashed: the abundances at the beginning of the \texttt{r256e1} simulation. Solid: after $\sim$8 hours, close to the end of the convective phase.}\label{fig:abund}
\end{figure}

\begin{figure}
\centering
\footnotesize
\includegraphics[trim={0cm 0 0cm 0cm},clip,width=0.48\textwidth]{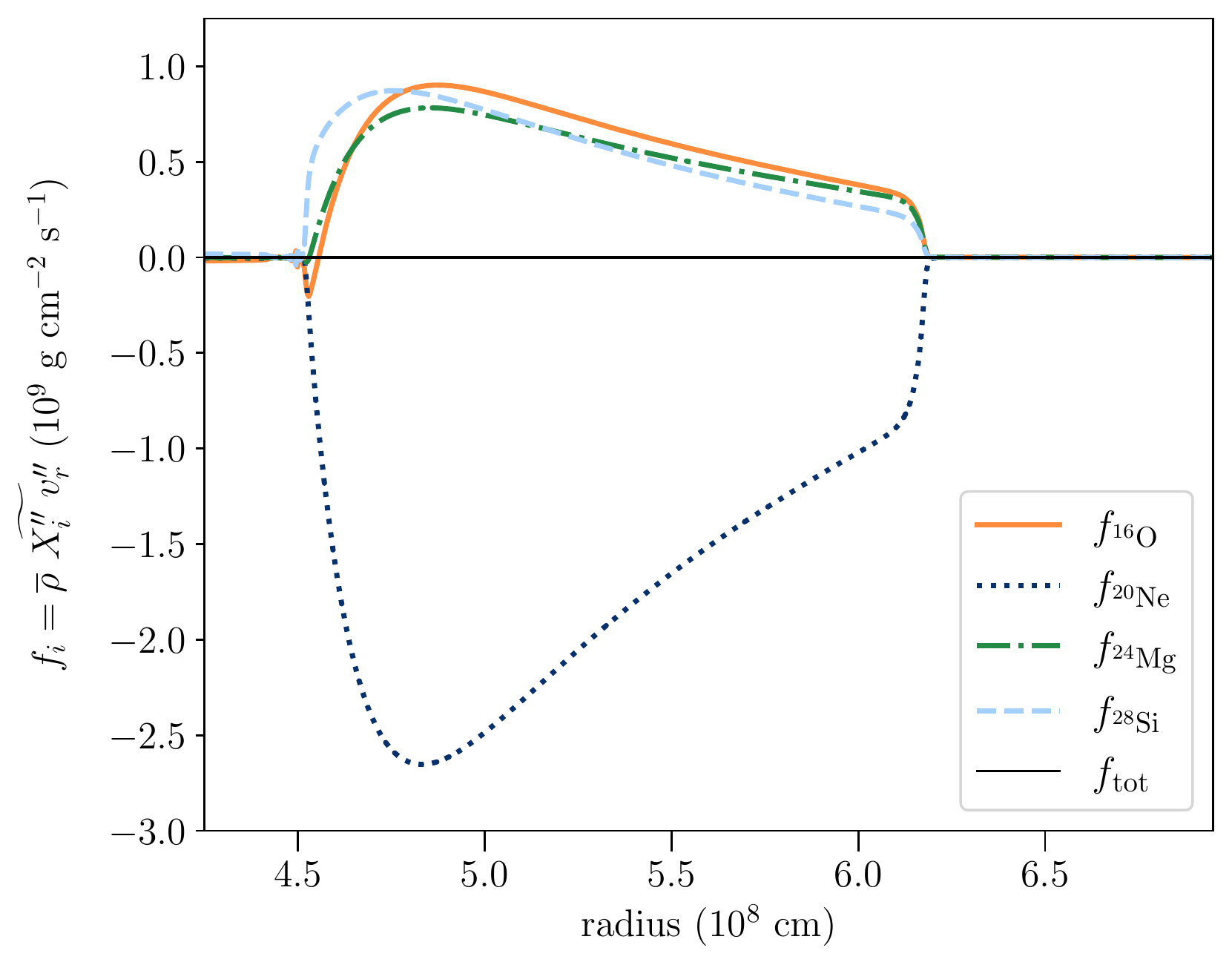}
\caption{Turbulent flux profiles of $^{16}$O, $^{20}$Ne, $^{24}$Mg, $^{28}$Si as function of the stellar radius, from the \texttt{r1024e1} simulation, averaged over the entrainment regime (3 convective turnovers), and defined as $f_\text{i}=\overline{\rho}\ \widetilde{X''_\text{i} v''_\text{r}}$. The thin black line is the sum of the flux profiles for all the 12 elements in the network.}\label{fig:flux}
\end{figure}

\begin{figure}
\centering
\footnotesize
\includegraphics[trim={0cm 0 0cm 0cm},clip,width=0.48\textwidth]{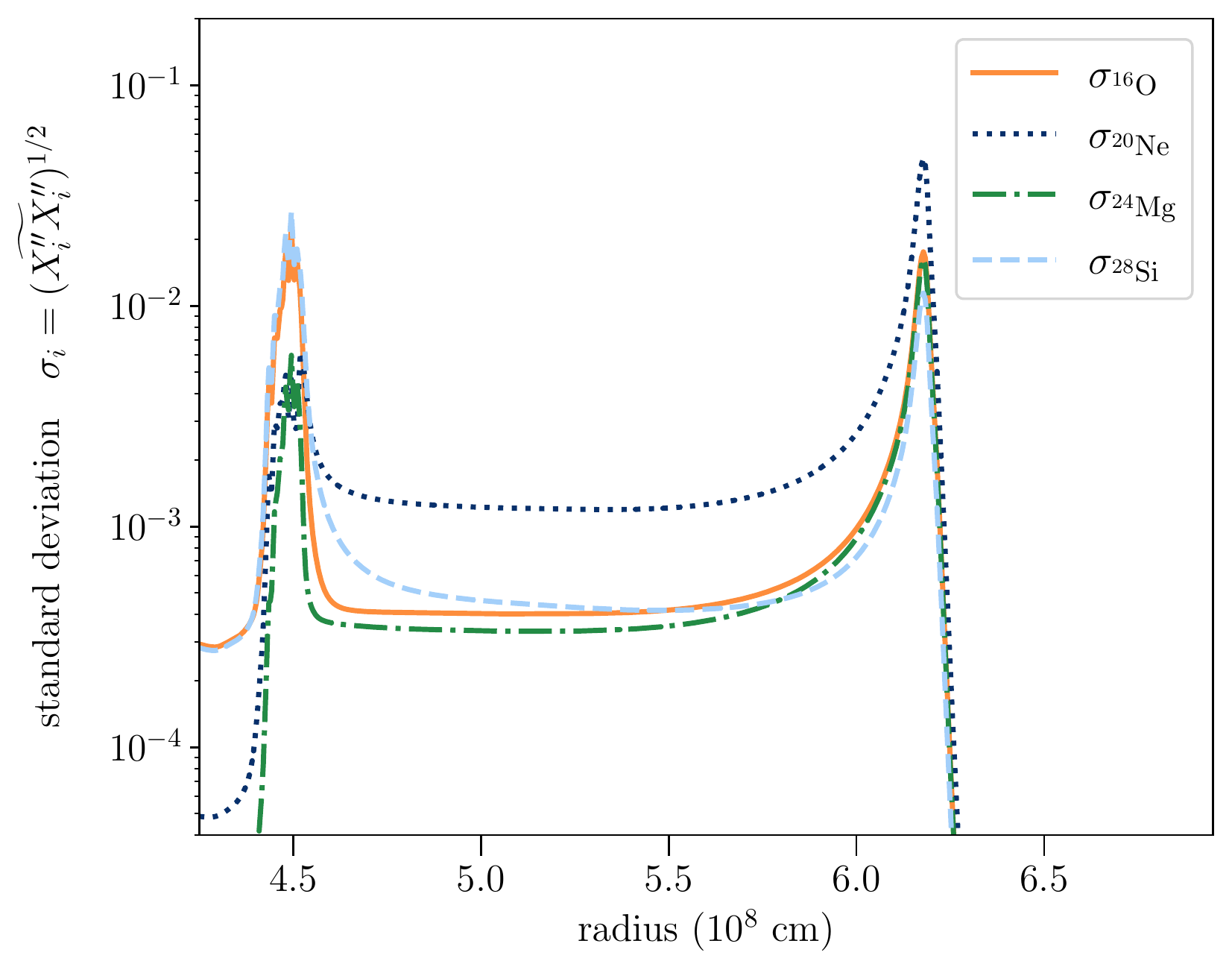}
\caption{Same as Fig.~\ref{fig:flux}, but standard deviation profiles of the mass fraction for $^{16}$O, $^{20}$Ne, $^{24}$Mg, $^{28}$Si, defined as $\sigma_\text{i}=(\widetilde{X''_\text{i} X''_\text{i}})^{1/2}$.}\label{fig:var}
\end{figure}

\begin{figure}
\centering
\footnotesize
\includegraphics[trim={0cm 0 0cm 0cm},clip,width=0.48\textwidth]{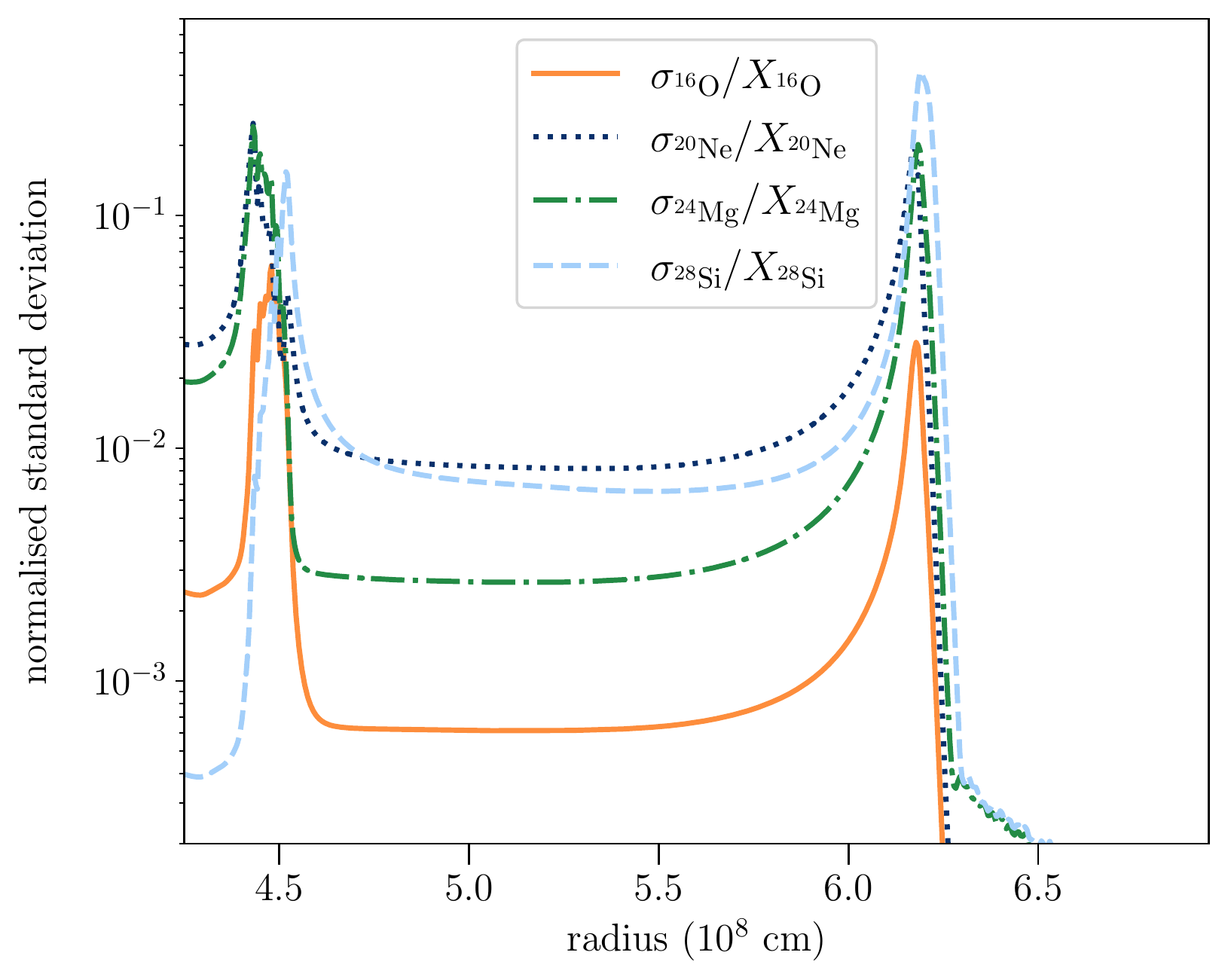}
\caption{Same as Fig.~\ref{fig:var}, but normalised standard deviation profiles for $^{16}$O, $^{20}$Ne, $^{24}$Mg, $^{28}$Si, defined as $\sigma_\text{i}/\overline{X}_\text{i}$.}\label{fig:varnorm}
\end{figure}

The simulations we present here have been produced employing a nuclear burning routine with an explicit list of isotopes to generate energy and drive convection. Making use of this routine, it is possible to study the time evolution of the chemical abundances and their distribution in the different layers of the simulations. We show in Fig.~\ref{fig:abund} the initial (dashed lines) and final (solid lines) mass fraction profiles from simulation \texttt{r256e1} for the most important isotopes involved in neon burning. At the beginning of the simulation the convective zone, identified by the central plateaus in the abundance profiles, is limited to the region between 4.5 and $5.8\times10^8$ cm, while towards the end it has almost doubled in size, as we have already seen from Fig.~\ref{fig:4pan}. $^{20}$Ne has been almost completely consumed in the convective zone via the reactions $^{20}$Ne$(\gamma,\alpha)^{16}$O and $^{20}$Ne$(\alpha,\gamma)^{24}$Mg, while $^{16}$O and $^{28}$Si have been produced as a result, as well as some $^{24}$Mg that has been partially burnt to produce silicon according to $^{24}$Mg$(\alpha,\gamma)^{28}$Si.
\begin{figure}
\centering
\footnotesize
\includegraphics[trim={0cm 0 0cm 0cm},clip,width=0.48\textwidth]{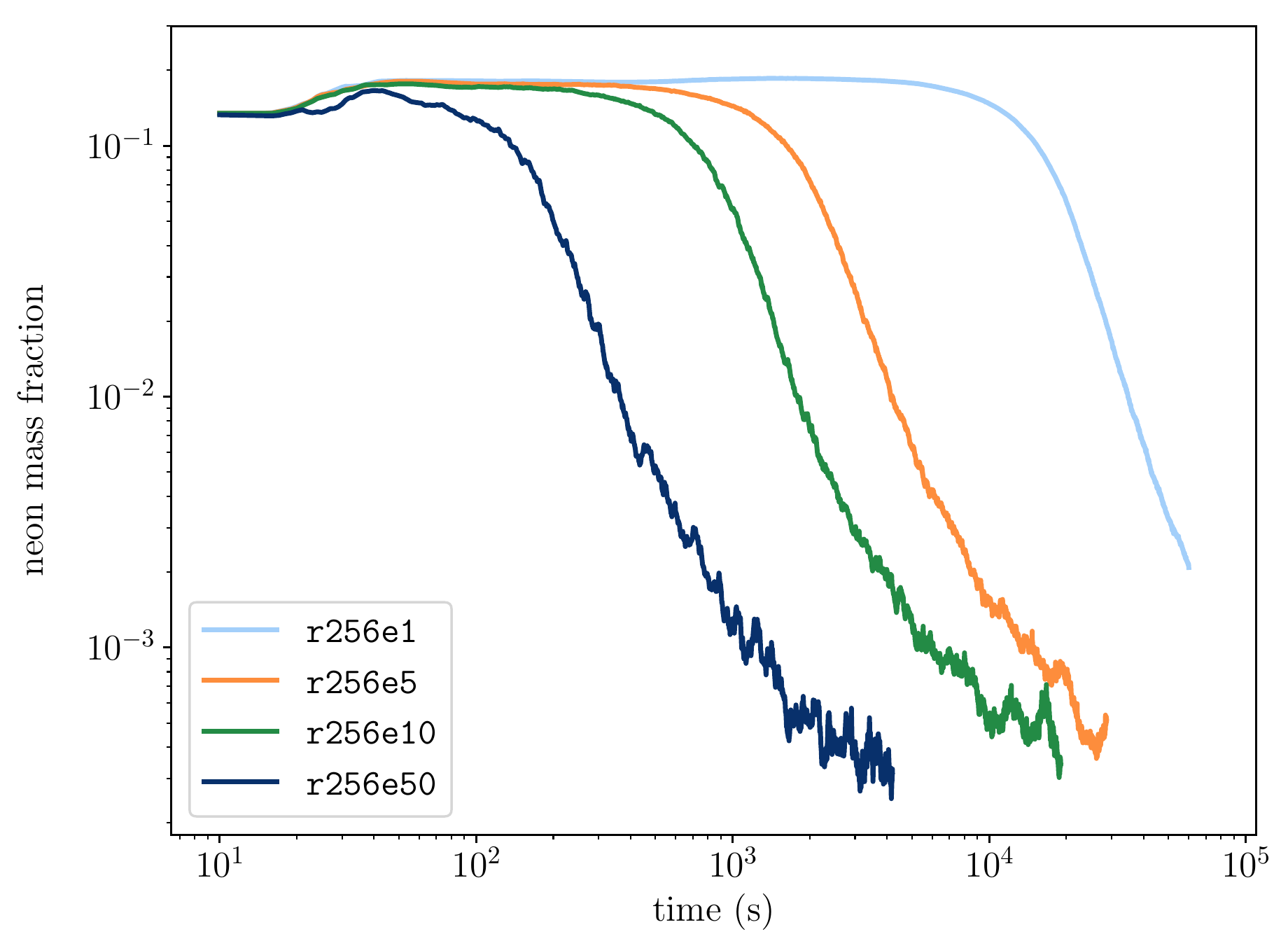}
\caption{Time evolution of the $^{20}$Ne mass fraction in the bulk of the convective zone, for the four different simulations \texttt{r256e1}, \texttt{r256e5}, \texttt{r256e10}, \texttt{r256e50}, each with a different boosting factor.}\label{fig:neon}
\end{figure}
\\Another way of tracking the neon consumption is to look at the time evolution of the neon abundance inside the convective zone, as we show in Fig.~\ref{fig:neon}. The plot shows the four simulations \texttt{r256e1}, \texttt{r256e5}, \texttt{r256e10}, \texttt{r256e50} with different boosting factors. During the initial transient phase (up to $\sim$100 s), some fluctuations in the neon abundance come from the initial propagation of plumes and eddies through the convection region, linked with the entrainment of some neon-rich material, and we can see that this trend is the same for all simulations except \texttt{r256e50}, where some neon is already being burnt due to the high energy boosting. After this phase, simulations consume neon on a different timescale but all with a very similar trend: this is an additional confirmation of the fact that the boosting in luminosity affects mainly the simulation timescale.\\
The chemical abundances can be also studied with a mean-field statistical analysis. We use here the Reynolds-Averaged Navier-Stokes (RANS) framework developed for hydrodynamic simulations in spherical geometry by \cite{2014arXiv1401.5176M}, making use of the dedicated open-source code \texttt{ransX}\footnote{\url{https://github.com/mmicromegas/ransX}}. We refer to \cite{2014arXiv1401.5176M, 2018M} for definitions and implementation. The RANS framework includes two types of averaging, a time averaging and an angular averaging. We will indicate here the Reynolds average (time and angular average) of a quantity $q$ on a spherical shell at radius $r$ as:
\begin{equation}
\overline{q}(r)=\dfrac{1}{T\Delta\Omega}\int_0^T \int_{\Delta\Omega} q(r,\theta,\phi,t)\ \mathrm{d} t\ \mathrm{d}\Omega
\end{equation}
with $\mathrm{d}\Omega=\sin\theta\mathrm{d}\theta\mathrm{d}\phi$ the solid angle element, $T$ the time window, $\Delta\Omega$ the solid angle of the shell. The Favre average (density-weighted average) is defined as $\widetilde{q}=\overline{\rho q}/\overline{\rho}$, therefore the field decomposition has been done as $q=\overline{q}+q'$ and $q=\widetilde{q}+q''$ respectively, with $\overline{q},\widetilde{q}$ the means and $q',q''$ the fluctuations of the quantity $q$ \citep[see][]{2014arXiv1401.5176M, 2018M}.\\
Figure \ref{fig:flux} shows the radial profiles of the mean turbulent flux in the \texttt{r1024e1} simulation for $^{16}$O, $^{20}$Ne, $^{24}$Mg, $^{28}$Si, defined as $f_\text{i}=\overline{\rho}\ \widetilde{X''_\text{i} v''_\text{r}}$ for a species $i$. Positive and negative values in the flux represent upward and downward flows respectively. It is evident that the flux is dominated by downward transport of $^{20}$Ne, towards the bottom of the convective zone where the nuclear burning is taking place. On the other hand, $^{16}$O, $^{24}$Mg and $^{28}$Si, the ashes of the burning, are all transported upwards in a similar way, through the entire convective zone. Additionally, a small non-zero flux is present immediately below the convective zone, slightly positive for silicon and negative for oxygen and magnesium. This is due to the mixing that takes place at the lower convective boundary, and the fact that silicon is more abundant below the boundary so it is brought inside the convective zone, while oxygen and magnesium are produced and more abundant above the boundary so they are transported downwards. Finally, the thin black line in Fig.~\ref{fig:flux} represents the sum of the flux profiles for all the 12 elements included in our nuclear network: it is always equal to zero, confirming that the sum of the mass fractions is conserved in our simulations.
\\Finally, we show in Fig.~\ref{fig:var} the standard deviation profiles of the mass fraction for $^{16}$O, $^{20}$Ne, $^{24}$Mg, $^{28}$Si, defined as $\sigma_\text{i}=(\widetilde{X''_\text{i} X''_\text{i}})^{1/2}$ for a species $i$, and in Fig.~\ref{fig:varnorm} the standard deviation normalised by the Reynolds-averaged mass fraction of the isotopes $\sigma_\text{i}/\overline{X}_\text{i}$, presenting deviations as fractions of the mean values\footnote{Note that in previous \texttt{PROMPI} studies, such as \cite{2018M}, the variable $\sigma_\text{i}$ is used to indicate the variance, rather than the standard deviation.}.
The standard deviation represents the dispersion of the chemical composition as a function of the radius, therefore it can be seen as a way of measuring the departure from a perfect spherical symmetry, providing important information for the comparison between 1D and multi-D stellar models. The largest deviations of up to a few tens of percents (see Fig.\,\ref{fig:varnorm}) are found at the convective boundary locations.
These deviations are explained by the deformation of the boundary due to the convective flow plowing into it as well as entrainment and the interaction of different layers at the interface (see e.\,g. Fig.\,\ref{fig:sect}). Inside the convective zone, on the other hand, the mixing makes the composition more homogeneous and reduces the dispersion down to a percent or less. 
It is interesting to note that deviations are still present below the convective region due to the fluctuations generated by entrainment and internal gravity waves.
\\Overall, the magnitude of the standard deviation is quite small and we do not expect major deviation from spherical symmetry for nucleosynthesis in normal convective burning episodes. The situation, however, is expected to be different in more dynamic contexts, such as merging shells or in cases where fuel is ingested in an unusual burning region \citep[see][]{2018M,2020MNRAS.491..972A}.

\section{Conclusions}\label{sec:4}
In this paper, we have presented a set of 3D hydrodynamic simulations of a complete stellar burning phase, a neon-burning shell in a 20 M$_\odot$ star. The accuracy of the simulations has been enhanced by improvements in geometry and resolution, nuclear network and burning routine, and initial conditions. We show that results from our simulations may be analysed in terms of nucleosynthesis, studying the abundance evolution and stratification of the isotopes included in the nuclear network, and of hydrodynamics. For the latter, we have analysed the convective motions and tracked the convective boundary evolution, allowing observation of the growth of the convective zone and its death when fuel is exhausted. Studying CBM is also an excellent way of comparing 1D and multi-D stellar simulations, where results are often in disagreement. CBM in 1D models is subject to uncertainties and needs calibration; hydrodynamic models can provide this calibration, but only if started from correct initial conditions. This shows how the two approaches are mutually dependent, and a convergence of results can only be achieved by improving one with the other. \\
In previous works, 1D models sometimes include little CBM in the convective core but totally ignore any CBM in later phases, when convection is even stronger and its effects are more important. In other works, multi-D models are started from initial conditions with little to no CBM, always leading to a very strong entrainment that is completely in disagreement with the initial 1D model. With this work, we make a step forward towards the convergence of 3D to 1D stellar models (321D approach). Our 3D simulations of a burning shell, run continuously from early development to fuel exhaustion, show that a) entrainment in late-phase shells does not proceed indefinitely as previously supposed, engulfing the entire star, but it halts when fuel is exhausted and convection dies (see Fig.~\ref{fig:4pan}), and b) starting from initial conditions already including strong CBM, the resulting entrainment in 3D is much more in agreement with the 1D model (see Fig.~\ref{fig:laura}). In particular for this last point, our entrainment study produced a law that may be equally well applied to CBM in convective cores (large RiB, vertical line in  Fig.~\ref{fig:laura}) and to late-phase shells (small RiB, data points in  Fig.~\ref{fig:laura}). This law may finally close the gap between the 1D and 3D stellar models, usually in disagreement regarding the amount of CBM to be included. Our results show that significant CBM is required not only in the convective core, as the most recent 1D models are starting to include, but also in the late-phase convective shells. The presence of large CBM is also supported by asteroseismic observations \citep[e.g.][]{2020B,2021NatAs...5..715P}.\\
The work presented in this paper introduces exciting prospects for stellar modelling. We have shown that simulating an entire burning phase in more than one dimension is now possible using the right tools and enough computing resources. The next few years will inevitably feature more simulations of significant fractions of the stellar lifetime in multi-D. Since covering the entire stellar evolution will probably never be possible in more than one dimension, the 1D stellar model will remain the main tool for predicting and explaining stellar evolution. However, it is the interplay between 1D and 3D models that really pushes forwards our knowledge of stellar evolution, and we show here that an agreement in results between the two is possible.\\
We recall here that the range of applications of stellar modelling to other branches of Astrophysics is large and various. This includes the production of accurate progenitor models as initial conditions for supernova explosion studies \citep{2015MNRAS.448.2141M,2019ApJ...881...16Y,2021Natur.589...29B} with possible deviations from spherical symmetry, potentially solving the long-standing core-collapse supernova engine problems, but also comparison to asteroseismic measurements \citep{2021A,2021NatAs...5..715P}, analysis and implementation of magnetic fields and dynamo effects in stars \citep{10.1093/mnras/stab883, 2022L}, predictions on the nature of the different remnants (white dwarfs, neutron stars, black holes) with improvements to the final-initial mass relation \citep{2020MNRAS.496.1967K,2021MNRAS.503.4208S}, nucleosynthesis and galactic chemical evolution. Furthermore, the synergy of theoretical models and observations will help tackle today’s open problems of stellar Astrophysics, such as the red supergiant problem \citep{2009ARA&A..47...63S} and the black hole mass gap \citep{2021ApJ...912L..31W}. 

\section*{Acknowledgements}
RH acknowledges support from the World Premier International Research Centre Initiative (WPI Initiative), MEXT, Japan and the IReNA AccelNet Network of Networks (National Science Foundation, Grant No. OISE-1927130). WDA acknowledges support from the Theoretical Astrophysics Program (TAP) at the University of Arizona and Steward Observatory. CG has received funding from the European Research Council (ERC) under the European Union’s Horizon 2020 research and innovation program (Grant No. 833925). The University of Edinburgh is a charitable body, registered in Scotland, with Registration No. SC005336. CG, RH, and CM acknowledge ISSI, Bern, for its support in organising collaboration. This article is based upon work from the ChETEC COST Action (CA16117) and the European Union’s Horizon 2020 research and innovation programme (ChETEC-INFRA, Grant No. 101008324). The authors acknowledge the STFC DiRAC HPC Facility at Durham University, UK (Grants ST/P002293/1, ST/R002371/1, ST/R000832/1, ST/K00042X/1, ST/H008519/1, ST/ K00087X/1, and ST/K003267/1). 

\section*{Data availability}
The data underlying this article will be shared on reasonable request to the corresponding author.




\bibliographystyle{mnras}
\bibliography{article} 


\bsp	
\label{lastpage}
\end{document}